\documentstyle[preprint,eqsecnum,aps]{revtex}
\begin{document}
\draft
\preprint{}

\title {\bf Factorization and polarization in linearized gravity}
\vskip 0.5cm

\author{S. Y. Choi}
\address{Theory Group, KEK, Tsukuba, Ibaraki 305, Japan}

\author{J. S. Shim}
\address{Department of Physics, Hanyang University,
         Seoul 133-791, Korea}

\author{H. S. Song}
\address{Center for Theoretical Physics and Department of
         Physics,\\
         Seoul National University, Seoul 151-742, Korea}

\maketitle

\begin{abstract}
We investigate all the four-body graviton interaction processes:
$gX\rightarrow \gamma X$, $gX\rightarrow gX$, and $gg\rightarrow gg$
with $X$ as an elementary particle of spin less than two in
the context of linearized gravity except the spin-3/2 case.
We show explicitly that gravitational gauge invariance and Lorentz
invariance cause every four-body graviton scattering amplitude to be
factorized. We explore the implications of this factorization
property by investigating polarization effects through the covariant
density matrix formalism in each four-body graviton scattering
process.
\vskip 0.4cm
\noindent
PACS number(s) : 04.60.+n, 12.25.+e, 13.88.+e
\end{abstract}


\section{Introduction}
\label{sec:intro}

Among the four fundamental interactions in nature, the gravitational
interaction has not yet been successfully quantized.
But the challenge of combining the quantum principle with the
elegant theory of general relativity, based on general covariance,
has been made ceaselessly.
While the very small gravitational coupling constant
might reduce the importance of theoretical and experimental
investigation of quantum gravity, gravity becomes as strong as
the other forces near the Planck scale, and it is believed to be crucial
in a consistent description of the birth of the Universe according
to the Big Bang senario.
Furthermore, the successful unification of electromagnetic and weak
interactions in the standard model makes unavoidable the thought that
further unifications might be realized for all other fundamental
interactions.
Recent developments of supergravity\cite{SG} and superstring
theories\cite{SS} were inspired by the hope of constructing a
consistent unified quantum theory including gravity.
In all cases, any common aspect of gravity and other interactions
is very much worth exploring.

It has been established by several people\cite{FL} that
the Fierz-Pauli theory of a massless spin-2 particle in the Minkowski
flat space-time is inconsistent when coupled to matter and the only
consistent theory in the low frequency domain is Einstein's general
relativity. In the light of this aspect, we use Einstein's general
relativity as a correct effective gravitational theory at low energies
compared to the Planck scale. Since we are interested mainly in
the weak field limit, we perform the weak field expansion to get
the linearized gravitational Lagrangian.
After the expansion, ordinary quantum field theoretical
methods are applied to the linearized gravity to obtain the
graviton-graviton and graviton-matter vertices.
Several graviton interaction processes have been studied previously
\cite{Gupta,Vlad,Miro,Voronov}
in this framework.

The formidable complexity in vertices with more than three gravitons
might render conventional Feynman diagram techniques very much
inefficient. Recently we have shown, however, that all the tree-level
transition amplitudes of $ge\rightarrow\gamma e$\cite{Choi1}, elastic
graviton-scalar, graviton-electron, graviton-photon, and
graviton-graviton scattering processes\cite{Choi2,Choi3},
are completely factorized into a simple form
composed of a kinematic factor, QED-like Compton scattering form, and
another gauge invariant terms. The factorization property can be used
as a powerful tool to investigate the gravitational interactions and
the polarization effects. The factorization property in the linearized
gravity corresponds to a well-known fact in the standard field theory
\cite{Grose,Goebel,Dongpei,Brodsky}
that gauge symmetry and Lorentz invariance enable all the
lowest-order ampilitudes of four-particle interactions with an external
massless gauge boson to be always factorized into one factor depending
on the charge or the internal symmetry indices and the other depending
on the spin or polarization indices.
A natural question is whether all the four-body graviton interactions
exhibit the same factorization property or not.

In this paper, we investigate in a more extensive way the four-body
graviton interactions like $gX\rightarrow \gamma X$ and
$gX\rightarrow gX$ in the context of linearized gravity, where $X$ is
any kind of particles with spin less than 2 or graviton itself.
Even though we do not consider the spin-3/2 case in the present work,
we considerably extend our previous works\cite{Choi1,Choi2,Choi3} to
show the presence of the factorization property in the four-body
graviton interactions including the case with a massive vector boson
$W$ for $X$.
In addition, we investigate the polarization effects to explore the
implications of the factorization property.

The paper is organized as follows. In Sec.~II, we describe in detail
the derivation of the gravitational Lagrangian for the graviton
scattering process with matter, including graviton itself,
and present its expanded form through the Gupta procedure\cite{Gupta}
in the weak field limit. Factorization in the linearized gravity is
explained in analogy with that of the standard gauge field theories
in Sec.~III. Sec.~IV is devoted to investigating polarization effects
in these graviton scattering processes and to exploring the
implications from the factorization property to the polarization
effects. A brief summary and discussion are given in Sec.~V.
Every Feynman rule needed in the present work is listed in the
Appendix.


\section{Interaction Lagrangian}
\label{sec:Lagrangian}

In this section, we describe a general procedure to derive
the gravitational Lagrangian for a graviton scattering with a massive
scalar, a massive fermion and a massive vector boson in the presence
of the electromagnetic field.
Without loss of generality it can be assumed that all the massive
particles have the same mass denoted by $m$.

The natural starting point for the derivation is the standard
QED Lagrangian in the absence of gravity:
\begin{eqnarray}
{\cal L}_{QED}&=& (D_\mu\phi)^*(D^\mu\phi)-m^2(\phi^*\phi)
     +i\bar{\psi}\gamma^\mu D_\mu \psi -m \bar{\psi}\psi\nonumber\\
  &&-\frac{1}{2}(D_\mu W_\nu -D_\nu W_\mu)^*(D^\mu W^\nu-D^\nu W^\mu)
     +m^2 W_\mu ^* W^\mu\nonumber\\
  &&-ieW_\mu^* W_\nu F^{\mu\nu}-\frac{1}{4}F^{\mu\nu}F_{\mu\nu},
\end{eqnarray}
where $\phi$ is a scalar field, $\psi$ is a fermion field,
$W$ is a vector boson field, and $A$ is a photon field, with which
the field strength $F_{\mu\nu}$ and the covariant derivative $D_\mu$
are defined as
\begin{eqnarray}
F_{\mu\nu}=\partial_\mu A_\nu -\partial_\nu A_\mu ,\   \
D_\mu=\partial_\mu +ie A_\mu.
\end{eqnarray}

The gravitational Lagrangian ${\cal L}$ is then obtained by making
the QED Lagrangian in a general covariant form.
To begin with, we write down the general covariant gravitational
Lagrangian without any detailed description of derivation,
\begin{eqnarray}
{\cal L}&=&{\cal L}_g+{\cal L}_{gs}(A)+{\cal L}_{gf}(A)
               +{\cal L}_{gW}(A)+{\cal L}_{gA}\label{act},\\
{\cal L}_g&=&2\kappa^{-2}\sqrt{-g}R,\label{pureg}\\
{\cal L}_{gs}(A)&=&\sqrt{-g}[g^{\mu\nu}(D_\mu \phi)^*(D_\nu \phi)
                 -m^2 \phi^* \phi],\\
{\cal L}_{gf}(A)&=&\sqrt{-g}\left[\frac{i}{2}\left(\bar{\psi}
                  \gamma^\mu (\vec{\nabla}_\mu -ieA_\mu) \psi
                 - \bar{\psi}(\stackrel{\leftarrow}{\nabla}_\mu
                 +ieA_\mu) \gamma^\mu \psi\right)
                 - m\bar{\psi}\psi \right] \label{lea},\\
{\cal L}_{gW}(A)&=&-\frac{1}{2} \sqrt{-g}g^{\mu\nu}g^{\alpha\beta}
                   (D_\mu W_\alpha -D_\alpha W_\mu)^*
                   (D_\nu W_\beta -D_\beta W_\nu) \nonumber\\
                &&+\sqrt{-g}g^{\mu\nu} m^2 W_\mu ^* W_\nu
                 - ie \sqrt{-g}g^{\mu\nu}g^{\alpha\beta}
                    W_\mu^* W_\alpha F_{\nu\beta},\\
{\cal L}_{gA}&=&-\frac{1}{4} \sqrt{-g}g^{\mu\nu}g^{\alpha\beta}
                  F_{\mu\alpha}F_{\nu\beta},
\end{eqnarray}
where $\kappa=\sqrt{32 G_N}$ with the Newtonian constant $G_N$.
For the sake of discussion, the gravitational Lagrangian ${\cal L}$
has been separated into five parts, each of which describes an
independent process under consideration.
The Lagrangian ${\cal L}_g$ describes pure gravitational
interactions. ${\cal L}_{gs}(A)$,  ${\cal L}_{gf}(A)$ and
${\cal L}_{gW}(A)$ are for gravitational interactions of a massive
scalar $s$, a massive fermion $f$, and a massive vector boson
$W$ in the presence of the electromagnetic field,
respectively. The final Lagrangian ${\cal L}_{gA}$
is for gravitational interactions of the electromagnetic field.

Now let us describe in detail the derivation procedure of
the gravitational Lagrangian ${\cal L}$ in the weak field limit
and expand the Lagrangian around the flat Minkowski space to get
necessary interaction terms.
The flat space expansion of Eq.\ (\ref{act}) usually can be
carried out by the Gupta procedure\cite{Gupta}.
In the procedure one introduces a symmetric tensor field
$h_{\mu\nu}$ denoting the deviation of the  metric tensor
$g_{\mu\nu}$  from the flat space Minkowski metric tensor
$\eta_{\mu\nu}$=($+,-,-,-$):
\begin{eqnarray}
g_{\mu\nu}=\eta_{\mu\nu}+\kappa h_{\mu\nu}.\label{gg}
\end{eqnarray}
After the expansion any curved space geometrical object
is expressed as an infinite series in terms of $h_{\mu\nu}$.
For the present work, however, only the terms up to O($h^3$)
are needed and therefore every expanded Lagrangian will be
presented including the terms up to that order.

It is convenient to expand at first the contravariant metric
tensor $g^{\mu\nu}$ and the Affine connection
$\Gamma^{\lambda}_{\mu\nu}$, whose expanded forms are given up
to O($h^3$) by
\begin{eqnarray}
&&g^{\mu\nu}=\eta^{\mu\nu}-\kappa h^{\mu\nu}
    +\kappa^2 h^{\mu\lambda} h^\nu_\lambda
    -\kappa^3 h^{\mu\lambda} h_{\lambda\alpha} h^{\alpha\nu},\\
      \label{gmu}
&&g\equiv {\rm det}(g_{\mu\nu})\nonumber\\
&&\hskip 1cm  =-1 -\kappa h
       +\frac{1}{2}\kappa^2(h^\mu_\rho h^\rho_\mu-h^2)
       +\frac{1}{6}\kappa^3(-2h^\mu_\rho h^\rho_\gamma
         h^\gamma_\mu+3hh^\mu_\rho h^\rho_\mu-h^3),\\
      \label{geq}
&&\sqrt{-g}=1 +\frac{\kappa}{2}h+\frac{\kappa^2}{8}(h^2
       -2h^\mu_\rho h^\rho_\mu)
       +\frac{\kappa^3}{48}(h^3-6hh^\mu_\rho h^\rho_\mu
       +8h^\mu_\rho h^\rho_\gamma h^\gamma_\mu),\\
&&{\Gamma^\lambda}_{\mu\nu}\equiv\frac{1}{2}  g^{\lambda\sigma}
       (\partial _\mu g_{\sigma\nu}+\partial_\nu g_{\sigma\mu}
       -\partial_\sigma g_{\mu\nu})\nonumber\\
      &&\hskip 1cm =\frac{1}{2} \kappa(\eta^{\lambda\sigma}
       -\kappa h^{\lambda\sigma}+\kappa^2 h^{\lambda\alpha}
       h^\sigma_\alpha)
        (\partial_\mu h_{\sigma\nu}+\partial_\nu h_{\sigma\mu}
       -\partial_\sigma h_{\mu\nu}),
       \label{GG}
\end{eqnarray}
with the definition $h=h^\mu_\mu$.

Let us now consider the Lagrangian ${\cal L}_g$ for pure
gravitational interactions.
The scalar curvature in Eq.\ (\ref{pureg}) is defined in terms
of the Affine connection $\Gamma^{\lambda}_{\mu\nu}$ as follows,
\begin{eqnarray}
R=g^{\mu\nu}\left[\partial_\nu {\Gamma^\lambda}_{\mu\lambda}
     -\partial_\lambda {\Gamma^\lambda}_{\mu\nu}
     +{\Gamma^\tau}_{\mu\lambda} {\Gamma^\lambda}_{\tau\nu}
     -{\Gamma^\tau}_{\mu\nu}
      {\Gamma^\lambda}_{\tau\lambda}\right].\label{r}
\end{eqnarray}
Taking the de Donder gauge
$\partial_\alpha h^{\alpha}_\mu=\frac{1}{2}
\partial_\mu h$, the Lagrangian ${\cal L}_g$ can be
expanded\cite{Berends,DeWitt} around the flat Minkowski space
and then reduced to the form:
\begin{eqnarray}
{\cal L}_g &=&{\cal L}_g^0 +\kappa {\cal L}_g^1
        +\kappa^2 {\cal L}_g^2 +\cdots,
       \label{simple}\\
{\cal L}_g^0 &=&-\frac{1}{4}\partial_\mu h\partial^\mu h
        +\frac{1}{2}\partial_\mu
         h^{\sigma\nu}\partial^\mu h_{\sigma\nu}, \\
{\cal L}_g^1&=&\frac{1}{2} h^\alpha_\beta \partial^\mu
         h^\beta_\alpha \partial_\mu h
       -\frac{1}{2}h^\alpha_\beta \partial_\alpha h^\mu_\nu
         \partial^\beta h^\nu_\mu-h^\alpha_\beta
         \partial_\mu h^\nu_\alpha
         \partial^\mu h^\beta_\nu \nonumber\\
       &&+\frac{1}{4}h\partial^\beta h^\mu_\nu \partial_\beta
         h^\nu_\mu+h^\beta_\mu \partial_\nu h^\alpha_\beta
         \partial^\mu h^\nu_\alpha
        -\frac{1}{8} h \partial^\nu h \partial_\nu h, \\
{\cal L}_g^2&=&\frac{1}{16}h^2 \partial_\mu h^{\alpha\beta}
         \partial^\mu h_{\alpha\beta}
        +h^\lambda_\mu h^\nu_\beta
         \partial_\lambda h^{\alpha\beta}\partial^\mu
         h_{\alpha\nu}-\frac{1}{8}h^{\mu\nu}h_{\mu\nu}
         \partial_\lambda h^{\alpha\beta}
         \partial^\lambda h_{\alpha\beta} \nonumber\\
       &&-2 h^{\lambda\nu} h_{\mu\nu}
         \partial_\lambda h^{\alpha\beta}\partial_\alpha
         h^\mu_\beta+\frac{1}{2} hh^\lambda_\mu
         \partial_\lambda h^{\alpha\beta}\partial_\alpha
         h^\mu_\beta -\frac{1}{2} hh^\mu_\beta
         \partial_\lambda h^{\alpha\beta}\partial^\lambda
          h_{\alpha\mu} \nonumber\\
       &&+h^\nu_\beta h^\mu_\nu
         \partial_\lambda h^{\alpha\beta}\partial^\lambda
         h_{\alpha\mu}
        -\frac{1}{2} h_{\alpha\beta} h^{\mu\nu}
         \partial_\lambda h^{\alpha\beta}
         \partial^\lambda h_{\mu\nu}
        +\frac{1}{2}h_\alpha^\mu h_\beta^\nu
         \partial_\lambda h^{\alpha\beta}
         \partial^\lambda h_{\mu\nu}\nonumber\\
       &&-\frac{1}{4} hh^\lambda_\mu
         \partial_\lambda h^{\alpha\beta}\partial^\mu
         h_{\alpha\beta}
        +\frac{1}{2}h^{\lambda\nu} h_{\mu\nu}
         \partial_\lambda h^{\alpha\beta}\partial^\mu
         h_{\alpha\beta} - h^\lambda_\beta h^\nu_\mu
         \partial_\lambda h^{\alpha\beta}\partial^\mu
         h_{\alpha\nu}\nonumber\\
       &&+\frac{1}{4}h h^\mu_\beta
         \partial_\lambda h\partial^\lambda h^\beta_\mu
        -\frac{1}{2}h^{\mu\nu}h_{\nu\beta}
         \partial_\lambda h\partial^\lambda h^\beta_\mu
        +\frac{1}{2} h^{\mu\nu}h_{\nu\beta}
         \partial_\lambda h\partial^\beta h^\lambda_\mu
         \nonumber\\
       &&-\frac{1}{4} h^{\mu\nu}h_{\mu\nu}
         \partial_\lambda h^{\alpha\beta}\partial_\alpha
         h_\beta^\lambda
        -\frac{1}{32}h^2\partial_\lambda h \partial^\lambda h
        +\frac{1}{8}h^{\mu\nu}h_{\mu\nu}
         \partial_\lambda h\partial^\lambda h.
\end{eqnarray}
We emphasize that  ${\cal L}^1_g$ and ${\cal L}^2_g$ have been
proved to be of the most compact form by a computer program
\cite{Jungil}.
While the difference is only a total derivative, the Lagrangian
(\ref{simple}) is much simpler than that of
Refs.~\cite{Berends,DeWitt}.
The gravitational Lagrangian ${\cal L}_{gs}(A)$ of a scalar in
the presence of the electromagnetic field can be similarly
expanded:
\begin{eqnarray}
&&{\cal L}_{gs}(A)={\cal L}_{gs}^0 +\kappa{\cal L}_{gs}^1
         +\kappa^2{\cal L}_{gs}^2+\cdots , \\
&&{\cal L}_{gs}^0=(D^\mu \phi)^* (D_\mu\phi)-m^2(\phi^*\phi), \\
&&{\cal L}_{gs}^1=\frac{1}{2}h {\cal L}_{gs}^0
        -h^{\mu\nu} (D_\mu\phi)^*(D_\nu\phi), \\
&&{\cal L}_{gs}^2=\frac{1}{8}(h^2 -2h^{\alpha\beta}h_{\alpha\beta})
           {\cal L}_{gs}^0
       +(h^\mu_\alpha h^{\alpha\nu} -\frac{1}{2}h h^{\mu\nu})
         (D_\mu\phi)^*(D_\nu\phi).
\end{eqnarray}

Let us now consider the gravitational Lagrangian of a fermion.
In the absence of gravity a free fermion is described by the
Lagrangian
\begin{eqnarray}
{\cal L}_f=\frac{i}{2}\left[\bar{\psi}\gamma^\mu\partial_\mu \psi
          -\partial^\mu\bar{\psi} \gamma_\mu \psi\right]
          -m\bar{\psi}\psi.
\end{eqnarray}
Incidentally, the fermionic Lagrangian ${\cal L}_f$ deserves a
special treatment when it is converted into a general covariant
form. Mathematically, this is because the tensor representations
of the GL(4) of general linear $4\times 4$ matrices behave like
tensors under the subgroup of Lorentz transformations,
but there is no representation of GL(4), or even representations
up to a sign, which behaves like spinor under the Lorentz subgroup.
One approach to incorporate spinors into general relativity is the
tetrad formalism\cite{Weinberg3}, which will be briefly described
below.

The formalism utilizes the fact that the equivalence principle
guarantees the introduction of a locally inertial coordinate
system ${y^m_P}$ at each space-time point $P$. In the case,
the metric tensor $g_{\mu\nu}$ is expressed as
\begin{eqnarray}
g_{\mu\nu}(x)=\eta_{mn} e^m_\mu(x) e^n_\nu(x),\label{v2}
\end{eqnarray}
where the tetrad or vierbein $e^m_\mu(x)$ is defined as a
coordinate derivative of $y^m_P$ as
\begin{eqnarray}
e^m_\mu(x)\equiv{\left[\frac{\partial y^m_P(x)}{\partial
                x^\mu}\right]}_{x=P}.
\end{eqnarray}
For the sake of discussion  a different type of vierbeins
$e^\nu_n$ is introduced with $m$ index lowered to $n$ with the
Minkowski metric tensor $\eta_{mn}$ and also with $\mu$ index
uppered to $\nu$ with the metric tensor $g^{\mu\nu}$;
\begin{eqnarray}
e^\nu_n\equiv \eta_{mn} g^{\mu\nu} e^m_\mu(x).
\end{eqnarray}
Eq.\ (\ref{v2}) shows that the vierbein ${e^\mu}_m$ is
nothing but the inverse of the vierbein ${e^m} _\mu$ such that
\begin{eqnarray}
\delta^\mu_\nu=e^\mu_m e^m_\nu,\ \
\delta^m_n=e^m_\mu e^\mu_n.
\end{eqnarray}

Another requirement from the equivalence principle is that the
special relativity should apply in locally inertial frames, i.e.,
should preserve Lorentz invariance locally.
As a way to accomplish the requirement  a new covariant derivative
is introduced;
\begin{eqnarray}
\nabla_m \equiv e^\mu_m (\partial_\mu +iw_\mu).
\end{eqnarray}
Then the locally Lorentz invariant gravitational Lagrangian of a
fermion is obtained as
\begin{eqnarray}
{\cal L}_{gf}=\frac{i}{2}\bar{\psi}\gamma^p e^\mu_p(\partial_\mu
        +iw_\mu )\psi+{\rm h.c.}-m\bar{\psi}\psi  ,
\end{eqnarray}
where the field connection $w_\mu(x)$ is expressed in terms of
vierbeins as
\begin{eqnarray}
w_\mu(x)=\frac{1}{4}\sigma^{mn}
     \left[e^\nu_m(\partial_\mu e_{n\nu}-\partial_\nu e_{n\mu})
    +\frac{1}{2}  e^\rho_m e^\sigma_n
     (\partial_\sigma e_{l\rho}-\partial_\rho e_{l\sigma})
     e^l_\mu-(m\leftrightarrow n)\right],
\end{eqnarray}
with $\sigma^{mn}=i[\gamma^m , \gamma^n]/2$ with the Dirac
matrices $\gamma^m$.
It can be now shown that the general covariant and
U(1)$_{\rm EM}$-invariant Lagrangian ${\cal L}_f (A)$ of
a fermion is
\begin{eqnarray}
{\cal L}_{gf}(A)= \sqrt{-g}\left[\frac{i}{2}
         \left\{\bar{\psi}\gamma^\mu
         (\vec{\nabla}_\mu-ieA_\mu)\psi
        -\bar{\psi}(\stackrel{\leftarrow}{\nabla}_\mu+ieA_\mu)
         \gamma^\mu\psi\right\}-m\bar{\psi}\psi\right] ,
\end{eqnarray}
with the notations
\begin{eqnarray}
\gamma^\mu=\gamma^p e^\mu_p,\ \
\vec{\nabla}_\mu \psi=\partial_\mu \psi+i w_\mu \psi ,\ \
\bar{\psi}{\stackrel{\leftarrow}{\nabla}}_\mu
    =\partial_\mu \bar{\psi}-i\bar{\psi} w_\mu.
\end{eqnarray}
In order to expand the Lagrangian around the flat Minkowski space
we first need to expand the vierbein $e^m_\mu$\cite{Woodard},
which is given by
\begin{eqnarray}
e^m_\mu=\delta^m_\mu +\frac{\kappa}{2} h^m_\mu
 -\frac{\kappa^2}{8}h^m_\nu\delta^\nu_nh^n_\mu+{\rm O}(\kappa^3).
\end{eqnarray}
The resulting Lagrangian ${\cal L}_{gf}(A)$ \cite{Voronov} is of
the following form;
\begin{eqnarray}
&&{\cal L}_{gf}(A)={\cal L}_{gf}^0
     +e\bar{\psi}\gamma^\mu \psi A_\mu +\kappa {\cal L}_{gf}^1
     +\frac{1}{2}\kappa e(h^\alpha_\alpha \eta_{\mu\nu}-h_{\mu\nu})
      \bar{\psi}\gamma^\mu\psi A^\nu+\kappa^2{\cal L}_{gf}^2
     +\cdots, \\
&&{\cal L}_{gf}^0=\frac{i}{2}[\bar{\psi}\gamma^\mu
     \partial_\mu \psi-\partial_\mu \bar{\psi}\gamma^\mu \psi]
    -m\bar{\psi}\psi ,\\
&&{\cal L}_{gf}^1=\frac{1}{2}  h {\cal L}_{gf}^0 -
     \frac{i}{4}h_{\mu\nu}[\bar{\psi}\gamma^\mu \partial^\nu \psi
    -\partial^\nu \bar{\psi}\gamma^\mu \psi],\\
&&{\cal L}_{gf}^2=\frac{1}{8} (h^2-2h^\alpha_\beta h^\beta_\alpha)
    {\cal L}_{gf}^0+\frac{i}{16}(3h_\nu^\alpha h_{\mu\alpha}
    -2h h_{\nu\mu} (\bar{\psi}\gamma^\mu \partial^\nu \psi
    -\partial^\nu \bar{\psi}\gamma^\mu \psi) \nonumber\\
&&\hskip 1cm+\frac{i}{16}[h^\nu_\alpha \partial^\alpha h^\mu_\nu
    -h^\mu_\nu \partial^\alpha h^\nu_\alpha]
     (\bar{\psi}\gamma_\mu \psi)
    +\frac{i}{32}[h^{\tau\alpha} \partial^\mu h^\nu_\alpha
    -h^{\nu\alpha} \partial^\mu h^\tau_\alpha]
     (\bar{\psi}\gamma_\mu \gamma_\nu \gamma_\tau\psi),
\end{eqnarray}
where $\gamma^\mu$ are the ordinary Dirac matrices and from now
on every Greek index refers to the flat Minkowskian space-time.

As in the scalar case, it is also possible to derive and expand
the gravitational Lagrangian for a vector boson in the presence of
the electromagnetic field;
\begin{eqnarray}
&&{\cal L}_{gW}(A)={\cal L}_{gW}^0 +\kappa {\cal L}_{gW}^1
                 +\kappa^2 {\cal L}_{gW}^2 \cdots, \\
&&{\cal L}_{gW}^0=-\frac{1}{2}(D_\mu W_\nu -D_\nu W_\mu)^*
              (D^\mu W^\nu -D^\nu W^\mu)
              +m^2 W_\mu^* W^\mu -ieW_\mu^* W_\nu F^{\mu\nu}, \\
&&{\cal L}_{gW}^1=h^{\mu\nu}\left[(D_\mu W_\alpha
              -D_\alpha W_\mu)^*
              (D_\nu W^\alpha -D^\alpha W_\nu)
              -m^2 W_\mu ^* W_\nu\right]\nonumber\\
              &&\hskip 1cm+\frac{1}{2}h {\cal L}_{gW}^0
              +ie[\eta^{\mu\nu}h^{\alpha\beta}
              +\eta^{\alpha\beta} h^{\mu\nu}]W_\mu^* W_\alpha
               F_{\nu\beta},\\
&&{\cal L}_{gW}^2=\frac{1}{8}(h^2-2h^{\mu\nu}h_{\mu\nu})
              {\cal L}_{gW}^0\nonumber\\
              &&\hskip 1cm+\frac{1}{2}(h h^{\alpha\rho}
               -2h^\alpha_\lambda h^{\lambda\rho})
              \left[\eta^{\beta\sigma}(D_\alpha W_\beta
              -D_\beta W_\alpha)^*
               (D_\rho W_\sigma-D_\sigma W_\rho)
               -m^2 W_\alpha^* W_\rho\right]\nonumber\\
              &&\hskip 1cm-h^{\mu\nu} h^{\alpha\beta}
               \left[\frac{1}{2}
               (D_\mu W_\alpha -D_\alpha W_\mu)^*
               (D_\nu W_\beta-D_\beta W_\nu)
             +ie W_\mu^* W_\alpha F_{\nu\beta}\right] \nonumber\\
              &&\hskip 1cm-ie\left[\eta^{\mu\nu}(h^{\alpha\lambda}
              h^\beta_\lambda
               -\frac{1}{2}h h^{\alpha\beta})
               +\eta^{\alpha\beta}(h^{\mu\lambda}h^\nu_\lambda
               -\frac{1}{2}h h^{\mu\nu})\right]
                W_\mu^* W_\alpha F_{\nu\beta}.
\end{eqnarray}
Finally, the gravitational Lagrangian of the electromagnetic
field is shown to be expanded as
\begin{eqnarray}
&&{\cal L}_{gA}={\cal L}_{gA}^0 +\kappa {\cal L}_{gA}^1
          +\kappa^2 {\cal L}_{gA}^2 +\cdots ,\\
&&{\cal L}_{gA}^0=-\frac{1}{4} F^{\mu\nu}F_{\mu\nu},\\
&&{\cal L}_{gA}^1=\frac{1}{2}h^\tau_\nu F^{\mu\nu} F_{\mu\tau}
          +\frac{1}{2} h {\cal L}_{gA}^0, \\
&&{\cal L}_{gA}^2=\frac{1}{8}(h^2-2h^\nu_\mu h^\mu_\nu)
            {\cal L}_{gA}^0 \nonumber\\
        &&\hskip 1cm+\frac{1}{4} F_{\alpha\beta}F_{\rho\sigma}
          \left[h h^{\alpha\rho}\eta^{\beta\sigma}
        -2h^\alpha_\mu h^{\mu\rho} \eta^{\beta\sigma}
          -h^{\alpha\rho} h^{\beta\sigma}\right].
\end{eqnarray}

To summarize, we have described in detail how to derive the
general covariant Lagrangian for gravitational interactions with
a scalar, a fermion and a vector boson in the presence of the
electromagnetic field. Then we have expanded the Lagrangian around
the flat Minkowski space through the Gupta procedure.
The expanded Lagrangian is not only Lorentz invariant at any order
of $h$ but also invariant under the transformation;
\begin{eqnarray}
h^{\mu\nu}\rightarrow h^{\mu\nu}+\partial^\mu X^\nu
         +\partial^\nu X^\mu,
\end{eqnarray}
with an arbitrary non-singular function $X^\mu$.
The latter invariance will be called gravitational gauge
invariance in the present work. It is now rather straightforward
to obtain all the Feynman rules of propagators and vertices up
to O($\kappa^2$). We present all the Feynman rules needed
in the present work in the Appendix.


\section{Factorization}
\label{sec:Factorization}

In the standard gauge theory every four-body Born amplitude with
a massless gauge boson as an external particle has been well
known to be factorizable\cite{Grose,Goebel,Dongpei,Brodsky} into
one factor which depends only on charge or other internal-symmetry
indices and the other factor which depends on spin or polarization
indices.

In this section we show that gravitational gauge invariance and
Lorentz invariance in the linearized gravity force all the
transition amplitudes of four-body graviton interactions to be
factorized as well.

First of all let us explain factorization in a (non-)Abelian
gauge theory following the procedure by Ref. \cite{Goebel}.
The crucial point is that any amplitude with an incoming gauge
boson is always arranged as a sum of terms of which each one
consists of three distinctive parts - a charge factor $A_i$,
a polarization-dependent part $B_i$, and a propagator $C_i$,
\begin{eqnarray}
{\cal M}=\sum^N_{i=1}\frac{A_i B_i}{C_i}.
\end{eqnarray}
Then each group factor is summed up to vanish;
\begin{eqnarray}
\sum^N_{i=1} A_i=\sum^N_{i=1} B_i=\sum^N_{i=1} C_i=0,
\label{412}
\end{eqnarray}
due to charge conservation (gauge invariance), energy-momentum
conservation (Lorentz invariance)
\begin{eqnarray}
\sum^N_{i=1}\delta_ip_i=k,
\end{eqnarray}
and transversality ($k\cdot\epsilon=0$), where $\epsilon^\mu$ and
$k^\mu$ are the polarization vector and 4-momentum of the massless
gauge boson, respectively, and $\delta_i=1(-1)$ is for an outgoing
(incoming) particle.
Every amplitude ${\cal M}$ for $N=3$ is then written in a
factorized form as
\begin{eqnarray}
{\cal M}=-\frac{C_1 C_2}{C_3}
          \left(\frac{A_1}{C_1}-\frac{A_2}{C_2}\right)
          \left(\frac{B_1}{C_1}-\frac{B_2}{C_2}\right),
\label{414}
\end{eqnarray}
or in equivalent forms with the indices (1,2,3) permuted.
It is now clear that the expression (\ref{414}) exhibits
factorization of the transition amplitude into a charge-dependent
part and a polarization-dependent part.

As an example, let us consider the gluon-gluon elastic scattering
process $G^aG^c\rightarrow G^bG^d$ where the superscripts denotes
color indces.
The factorization theorem enables the transiton amplitude to be
factorized as \cite{Goebel,Dongpei}
\begin{eqnarray}
{\cal M}\Bigl(G^aG^c\rightarrow G^bG^d\Bigr)
  =
 \left(\frac{\alpha_s}{\alpha}\right)
 \frac{(p_1\cdot k_1)(p_1\cdot k_2)}{k_1\cdot k_2}
 \left[\frac{f^{ace}f^{bde}}{p_1\cdot k_1}
 -\frac{f^{ade}f^{bce}}{p_1\cdot k_2}\right]
 {\cal M}_{\gamma v},
 \label{Mgg}
\end{eqnarray}
where $k_1(p_1)$ are 4-momenta of the incident gluon $G^a(G^c)$,
$k_2(p_2)$ are 4-momenta of the final gluon $G^b(G^d)$, and
$v$ stands for a massless vector boson with a positive electric
charge. Here, $f^{abc}$ are the structure constants of the SU(3)
color-gauge group.  The amplitude ${\cal M}_{\gamma v}$, which is
of the same form as the Compton scattering amplitude of a charged
massless vector boson, is given by
\begin{eqnarray}
{\cal M}_{\gamma v}
        =e^2\left[\frac{B_1}{C_1}-\frac{B_2}{C_2}\right],
\end{eqnarray}
where
\begin{eqnarray}
&&B_1=-\epsilon^\mu_1\epsilon^{*\nu}_2
  \varepsilon^\alpha_1(p_1)\varepsilon^{*\beta}_2(p_2)
  [C_{\mu\alpha\delta}(k_1,p_1,-q_1)
  {C_{\nu\beta}}^\delta(-k_2,-p_2,q_1)
  +2p_1\cdot k_1(\eta_{\mu\nu}\eta_{\alpha\beta}
  - \eta_{\mu\beta}\eta_{\nu\alpha})], \nonumber\\
&&B_2=\epsilon^\mu_1\epsilon^{*\nu}_2
  \varepsilon^\alpha_1(p_1)\varepsilon^{*\beta}_2(p_2)
  [C_{\mu\beta\delta}(k_1,-p_2,q_2)
   {C_{\nu\alpha}}^\delta(-k_2,p_1,-q_2)
  -2p_1\cdot k_2(\eta_{\mu\nu}\eta_{\alpha\beta}
  -\eta_{\mu\alpha}\eta_{\nu\beta})], \nonumber\\
&&B_3=\epsilon^\mu_1\epsilon^{*\nu}_2
  \varepsilon^\alpha_1(p_1)\varepsilon^{*\beta}_2(p_2)
  [C_{\mu\nu\delta}(k_1,-k_2,-q_3)
  {C_{\alpha\beta}}^\delta(p_1,-p_2,q_3)
  -2k_1\cdot k_2(\eta_{\mu\alpha}\eta_{\nu\beta}
  -\eta_{\mu\beta}\eta_{\nu\alpha})],\nonumber\\
&&{ }   \label{Bi}\\
&&C_1=2(p_1\cdot k_1),\ \ C_2=-2(p_1\cdot k_2),
  \ \ C_3=-2(k_1\cdot k_2),
   \label{Ci}
\end{eqnarray}
with the definition
\begin{eqnarray}
C_{\lambda\mu\nu}(p,q,r)=(p-q)_\nu \eta_{\lambda\mu}
+(q-r)_\lambda\eta_{\mu\nu}+(r-p)_\mu\eta_{\nu\lambda}.
\end{eqnarray}
The transfered momenta $q_1$, $q_2$, and $q_3$ are given in
terms of external momenta by
\begin{eqnarray}
q_1=p_1+k_1,\ \ g_2=p_1-k_2,\ \ q_3=k_1-k_2.
\end{eqnarray}
It is a simple matter to determine the $A_i$ factors for the
process $G^aG^c\rightarrow G^bG^d$ by the use of the Jacobi
identity of the structure functions;
\begin{eqnarray}
A_1=f^{ace}f^{dbe}, \ \ A_2=f^{ade}f^{bce},
\ \ A_3=f^{abe}f^{cde}.
\end{eqnarray}

We emphasize that the factorization in an ordinary gauge field
theory stems from guage invariance and Lorentz invariance of the
theory. Since the linearized gravity is gravitational guage
invariant as well as Lorentz invariant, it is expected to have
a similar factorization property in the linearized
gravity. In the present work we show explicitly that every
amplitude of a graviton scattering with a scalar, a fermion,
a vector boson or a graviton itself indeed exhibits such a
remarkable factorization.

In order to prove factorization in the linearized gravity, we
first note that gravitational guage invaiance guarantees the
decomposition\cite{Gross} of a graviton wave tensor
$\epsilon^{\mu\nu}(2\lambda)$ into a multiplication of two
spin-1 wave vectors,
\begin{eqnarray}
\epsilon^{\mu\nu}(2\lambda)=\epsilon^\mu(\lambda)
                            \epsilon^\nu(\lambda),
\label{415}
\end{eqnarray}
where the wave vector $\epsilon^\mu(\lambda)$ satisfies
\begin{eqnarray}
k\cdot\epsilon(\lambda)=0,\ \ \epsilon(\lambda)\cdot
      \epsilon(\lambda^\prime)
    =-\delta_{\lambda,-\lambda^\prime},
\end{eqnarray}
so that the wave tensor $\epsilon^{\mu\nu}(2\lambda)$ satisfies
\begin{eqnarray}
k_{\mu} \epsilon^{\mu\nu}(2\lambda)=\epsilon^{\mu\nu}(2\lambda)
        k_{\nu}=0,\ \
{\epsilon^\mu}_\mu(2\lambda)=0,
\end{eqnarray}
with the graviton 4-momentum $k$.

In order to show clearly the common features of four-body graviton
processes, we organize the presentation of our results in the
following. First, we introduce $X$ as a generic notation for a
scalar $s$, a fermion $f$, or a vector boson $W$.
Second, the $k_i$ ($i=1,2$) are for the 4-momenta of the incident
$g$ and the final $g(\gamma)$ in the process
$gX\rightarrow g(\gamma)X$, and $p_i$ ($i=1,2$)
for the 4-momenta of the initial $X$ and the final $X$,
respectively.
We note that there are four Feynman diagrams for every process
(see Figs.~2 and 3).
The last diagram in each figure set is a so-called contact term,
which can be always absorbed into the other parts.
We present only the results after absorbing the
contact term and rearranging the amplitude according to the
factorization theorem. Finally, we mention that all the processes
under consideration have the same set of the kinematical factors
$C_i$, denoting the $s$, $t$, and $u$ channel momentum transfers,
\begin{eqnarray}
C_1=2(p_1\cdot k_1),\ \ C_2=-2(p_1\cdot k_2),
   \ \ C_3=-2(k_1\cdot k_2),
\end{eqnarray}
which are the same as Eq.~(\ref{Ci}).
For simplicity, we will no longer write down this kinematical
set in the following.

\subsection{Graviton conversion into a photon}

In this subsection, we consider the process of a graviton
scattering off a particle $X$ for the photon production, where
$X$ can be a scalar $s$, a fermion $f$, or a vector boson $W$.
The graviton conversion into a photon can be considered as a
means\cite{Novaes} to detect a gravitational wave.
As mentioned before, $k_1$ and $\epsilon_1$ are the incident
graviton momentum and a wave vector for the graviton wave tensor,
while $k_2$ and $\epsilon_2$ are the final photon momentum and
the photon wave vector, repectively. $p_1(p_2)$ denote the
4-momentum of the incident(final) $X$ particle.

\subsubsection{\mbox{$gs\rightarrow \gamma s$}}

The simplest nontrivial process is the graviton scattering off
a scalar particle for a photon production
$gs\rightarrow\gamma s$.
The process is of order $e$ and $\kappa$ in both the gravitational
and the electromagnetic interactions and therefore the relevant
Lagrangian consists of four parts as
\begin{eqnarray}
{\cal L}^{\gamma s}_{I}={\cal L}_{gs}^0(A)
              +\kappa {\cal L}_{gs}^1(A)
              +{\cal L}_{gA}^0+\kappa {\cal L}_{gA}^1.
\end{eqnarray}
The Feynman diagrams are shown in Fig.~2, where the solid line is
for the scalar particle.
After absorbing the contact term denoted by the last diagram,
we obtain the resulting transition amplitude for the process
$gs\rightarrow\gamma s$ devided into three parts;
\begin{eqnarray}
{\cal M}_{gs\rightarrow\gamma s}={\cal M}^{\gamma s}_a
        +{\cal M}^{\gamma s}_b+{\cal M}^{\gamma s}_c,
\end{eqnarray}
\begin{eqnarray}
&&{\cal M}^{\gamma s}_a=e\kappa
       \frac{(p_1\cdot\epsilon_1)}{2(q^2_1-m^2)}
       \left[(q^2_1-m^2)(\epsilon_1\cdot \epsilon^*_2)
      -4(p_2\cdot \epsilon^*_2)(p_1\cdot \epsilon_1)\right],\\
&&{\cal M}^{\gamma s}_b=e\kappa
       \frac{(p_2\cdot \epsilon_1)}{2(q^2_2-m^2)}
       \left[(q^2_2-m^2)(\epsilon_1\cdot \epsilon^*_2)
      -4(p_2\cdot \epsilon_1)(p_1\cdot \epsilon^*_2)\right],\\
&&{\cal M}^{\gamma s}_c=e\kappa
       \frac{(k_2\cdot \epsilon_1)}{2q^2_3}
       \left[(q^2_1-q^2_2)(\epsilon_1\cdot \epsilon^*_2)
          +4(p_2\cdot\epsilon_1)(p_1\cdot\epsilon^*_2)
          -4(p_2\cdot\epsilon^*_2)(p_1\cdot\epsilon_1)\right].
\end{eqnarray}
After extracting the kinematical factors $C_i$, it is
straightforward to determine $A^{\gamma s}_i$ and $B^{\gamma s}_i$
($i=1,2,3)$ as
\begin{eqnarray}
A^{\gamma s}_1&=&e\kappa(\epsilon_1\cdot p_1),\ \
A^{\gamma s}_2 = -e\kappa(\epsilon_1\cdot p_2),\ \
A^{\gamma s}_3 = -e\kappa(\epsilon_1\cdot k_2), \\
B^{\gamma s}_1&=&(p_1\cdot k_1) (\epsilon_1\cdot\epsilon^*_2)
   -2(p_2\cdot\epsilon^*_2)( p_1\cdot\epsilon_1),\nonumber\\
B^{\gamma s}_2&=&(p_2\cdot k_1)(\epsilon_1\cdot\epsilon^*_2)
   +2(p_2\cdot\epsilon_1)( p_1\cdot\epsilon^*_2),\nonumber\\
B^{\gamma s}_3&=&-[(p_2+p_1)\cdot k_1](\epsilon_1\cdot\epsilon^*_2)
   -2(p_2\cdot\epsilon_1)(p_1\cdot\epsilon^*_2)
   +2(p_2\cdot\epsilon^*_2)(p_1\cdot\epsilon_1).
\end{eqnarray}
The transition amplitude is reduced to the factorized form
\begin{eqnarray}
{\cal M}_{gs \rightarrow \gamma s}=-e\kappa F
         \left[\frac{\epsilon_1\cdot p_1}{p_1\cdot k_1}
        -\frac{\epsilon_1\cdot p_2}{p_1\cdot k_2}\right]
         \left[(\epsilon_1\cdot\epsilon^*_2)
        -\frac{(\epsilon_1\cdot p_1)
         (\epsilon^*_2\cdot p_2)}{p_1\cdot k_1}
        +\frac{(\epsilon_1\cdot p_2)
         (\epsilon^*_2\cdot p_1)}{p_1\cdot k_2}
          \right],
\end{eqnarray}
where $\epsilon^\mu_1$ and $\epsilon^\mu_2$ are two wave vectors
for a graviton and a photon, respectively.
Here, the overall kinematical factor $F$ is
\begin{eqnarray}
F=\frac{(p_1\cdot k_1)(p_1\cdot k_2)}{(k_1\cdot k_2)}.
\end{eqnarray}
We note that the last factor is of the same form as the
scalar-Compton scattering amplitude.

Before proceeding further, let us note here that the introduction
of a manifestly gauge invariant four-vector $\tilde{\epsilon_i}$
($i=1,2$);
\begin{eqnarray}
\tilde{\epsilon_i}=\epsilon_i
    -\frac{(p_1\!\cdot\!\epsilon_i)}{(p_1\!\cdot\! k_i)}k_i,
 \label{fac3}
\end{eqnarray}
renders the expression of the scalar Compton scattering amplitude
${\cal M}_{\gamma s}$ greatly simplified;
\begin{eqnarray}
{\cal M}_{\gamma s}=2e^2
  \bigl(\tilde{\epsilon}_1\!\cdot\!\tilde{\epsilon}^*_2\bigr).
  \label{fac4}
\end{eqnarray}
Along with this simplification, gravitational gauge invariance
and graviton transversality render the transition amplitude
${\cal M}_{gs \rightarrow \gamma s}$ completely factorized into a
quite simple form;
\begin{eqnarray}
{\cal M}_{gs \rightarrow \gamma s}
   =\frac{\kappa}{2e}F
    \left(\frac{p_2\cdot\tilde{\epsilon}_1}{p_2\cdot k_1}\right)
    \left[{\cal M}_{\gamma s}\right].
 \label{fac2}
\end{eqnarray}
\subsubsection{\mbox{$gf\rightarrow\gamma f$}}

The next simplest case is the process $gf\rightarrow\gamma f$
with a graviton in place of the incident photon in the ordinary
fermion-Compton scattering process.
The relevant Lagrangian for the process $gf\rightarrow\gamma f$
is composed of four parts;
\begin{eqnarray}
{\cal L}^{\gamma f}_{I}
   ={\cal L}_{gf}^0(A)+\kappa {\cal L}_{gf}^1(A)
   +{\cal L}_{gA}^0 +\kappa {\cal L}_{gA}^1.
\end{eqnarray}

The Feynman diagrams for the process can be drawn in the same
way as for the process $gs\rightarrow \gamma s$.
The only difference is that in the present case the solid line
in Fig.~2 is for a fermion.  After absorbing the contact term
denoted by the last diagram in Fig.~2, we obtain the transition
amplitude ${\cal M}_{gf \rightarrow \gamma f}$;
\begin{eqnarray}
{\cal M}_{gf\rightarrow\gamma f}&=&{\cal M}^{\gamma f}_a
   +{\cal M}^{\gamma f}_b
   +{\cal M}^{\gamma f}_c,
\end{eqnarray}
\begin{eqnarray}
&&{\cal M}^{\gamma f}_a=-\frac{e\kappa}{2}
     \frac{(\epsilon_1\cdot p_1)}{(q_1^2-m^2)}
     \bar{u}(p_2,s_2)\left[\not\!{\epsilon^*_2}
     (\not\!{q_1}+m)\!\not\!{\epsilon_1}\right]u(p_1,s_1),\\
&&{\cal M}^{\gamma f}_b
     =-\frac{e\kappa}{2}\frac{(\epsilon_1\cdot p_2)}
      {(q_2^2-m^2)} \bar{u}(p_2,s_2)\left[\not\!{\epsilon_1}
     (\not\!{q_2}+m)\!\not\!{\epsilon^*_2}\right]u(p_1,s_1),\\
&&{\cal M}^{\gamma f}_c
     =\frac{e\kappa}{q_3^2} (k_2\cdot \epsilon_1)
      \bar{u}(p_2,s_2)\bigl[(\epsilon_1\cdot\epsilon^*_2)
      \not\!{k_2}-(\epsilon_1\cdot k_2)\not\!{\epsilon^*_2}
      -(\epsilon^*_2\cdot k_1)\not\!{\epsilon_1}\bigr]u(p_1,s_1).
\end{eqnarray}
Now the factors $A_i$ and $B_i$ ($i=1,2,3$) for the process
$gf\rightarrow\gamma f$ are
\begin{eqnarray}
A^{\gamma f}_i&=&A^{\gamma s}_i \  \  (i=1,2,3),\\
B^{\gamma f}_1
      &=&-\frac{1}{2}\bar{u}(p_2,s_2)\left[\not\!{\epsilon_2}
         (\not\!{q_1}+m)\not\!\!{\epsilon_1}
          \right]u(p_1,s_1),\nonumber\\
B^{\gamma f}_2
      &=&\frac{1}{2}\bar{u}(p_2,s_2)\left[\not\!{\epsilon_1}
          (\not\!{q_2}+m)\not\!\!{\epsilon^*_2}
           \right]u(p_1,s_1),\nonumber\\
B^{\gamma f}_3&=&\bar{u}(p_2,s_2)\left[\not\!{\epsilon_1}
        (\epsilon^*_2\cdot k_1)
        +\not\!{\epsilon^*_2}(\epsilon_1\cdot k_2)
        -(\epsilon_1\cdot\epsilon_2)\not\!\!{k_2}\right]u(p_1,s_1).
\end{eqnarray}
As mentioned before the $C_i$ factors are the same as those for
the process $gs\rightarrow\gamma s$.
Consequently we are led to the factorized transition amplitude,
\begin{eqnarray}
{\cal M}_{gf\rightarrow \gamma f}&=&\frac{\kappa}{2e} F
      \left(\frac{p_2\cdot\tilde{\epsilon}_1}{p_2\cdot k_1}\right)
      \left[{\cal M}_{\gamma f}\right],
\end{eqnarray}
where the transition amplitude ${\cal M}_{\gamma f}$ is of the same
form  as the standard Compton scattering amplitude
\begin{eqnarray}
{\cal M}_{\gamma f}=-e^2\bar{u}(p_2,s_2)\left[\not\!{\epsilon^*_2}
   \frac{1}{\not\!{q_1}-m}\!\not\!{\epsilon_1}+\not\!{\epsilon_1}
   \frac{1}{\not\!{q_2}-m}\!\not\!{\epsilon^*_2}\right]u(p_1,s_1).
   \label{Mrf}
\end{eqnarray}
\subsubsection{\mbox{$gW\rightarrow\gamma W$}}

Since the process $gW\rightarrow\gamma W$ involves three vector
particles and one graviton, Feynman rules are complicated and as
a result the expression of the amplitude is complicated as well.
Without any new insight of the amplitude structure, any conventional
method will require a lot of time to calculate the cross section.
This formidable algebra can be avoided by a simple reorganization
of the amplitude due to the factorization as described below.

First we write down the relevant Lagrangian for the process
$gW\rightarrow \gamma W$, which consist of four parts;
\begin{eqnarray}
{\cal L}^{\gamma W}_{I}={\cal L}_{gW}^0(A)
            +\kappa {\cal L}_{gW}^1(A)
            +{\cal L}_{gA}^0 +\kappa {\cal L}_{gA}^1.
\end{eqnarray}
Through the same procedure as in the process
$gf\rightarrow\gamma f$ we find
\begin{eqnarray}
A^{\gamma W}_i&=&A^{\gamma s}_i\ \ (i=1,2,3),\\
B^{\gamma W}_1&=&\frac{1}{2} \left[B_1
         -m^2(\epsilon_1\cdot\varepsilon_1)
         (\epsilon^*_2\cdot\varepsilon^*_2)\right],\nonumber\\
B^{\gamma W}_2&=&\frac{1}{2}\left[B_2
         +m^2(\epsilon_1\cdot\varepsilon^*_2)
          (\epsilon^*_2\cdot\varepsilon_1)\right],\ \
B^{\gamma W}_3=\frac{B_3}{2},
\end{eqnarray}
where $B_i$ ($i=1,2,3$) are given in Eq.~(\ref{Bi}).
Then the amplitude reduces to the factorized form as
\begin{eqnarray}
{\cal M}_{gW\rightarrow \gamma W}&=&-e\kappa F
 \left(\frac{p_2\cdot\tilde{\epsilon}_1}{p_2\cdot k_1}\right)
 \left[(\tilde{\epsilon_1}\!\cdot\!\tilde{\epsilon^*_2})
       (\tilde{\varepsilon_1}\!\cdot\!\tilde{\varepsilon^*_2})
      \!-(k_1\!\cdot\! k_2)\!
 \left\{\frac{(\tilde{\epsilon_1}\!\cdot\!\tilde{\varepsilon_1})
       (\tilde{\epsilon^*_2}\!\cdot\!
        \tilde{\varepsilon^*_2})}{p_1\!\cdot\! k_1}
       -\frac{(\tilde{\epsilon_1}\!\cdot\!\tilde{\varepsilon^*_2})
        (\tilde{\epsilon^*_2}\!\cdot\!
         \tilde{\varepsilon_1})}{p_1\!\cdot\! k_2}\right\}
         \right.\nonumber\\
   &&\left. \hskip 1.9cm-\frac{1}{p_1\!\cdot\! k_2}
        \{(\tilde{\epsilon^*_2}\!\cdot\!\tilde{\varepsilon_1})
        (p_2\!\cdot\!\tilde{\epsilon_1})
        (k_2\!\cdot\!\tilde{\varepsilon^*_2})
       +(\tilde{\epsilon_1}\!\cdot\!\tilde{\varepsilon^*_2})
        (p_2\!\cdot\!\tilde{\epsilon^*_2})
        (k_2\!\cdot\!\tilde{\varepsilon_1})\right.\nonumber\\
   &&\left. \hskip 1.9cm-(\tilde{\epsilon^*_2}
        \!\cdot\!\tilde{\varepsilon^*_2})
        (p_2\!\cdot\!\tilde{\epsilon_1})
        (k_2\!\cdot\!\tilde{\varepsilon_1})\}
       +\frac{1}{p_1\!\cdot\! k_1}
        (\tilde{\epsilon_1}\!\cdot\!\tilde{\varepsilon_1})
        (p_2\!\cdot\!\tilde{\epsilon^*_2})
        (k_2\!\cdot\!\tilde{\varepsilon^*_2})\right],
     \label{fac5}
\end{eqnarray}
where $p_1$($p_2$) and $\varepsilon_1$($\varepsilon_2$) are the
four-momentum and wave vector of the initial(final) vector boson,
respectively. The $\tilde{\varepsilon_i}$ $(i=1,2)$ is a
polarization vector defined in a manifestly gauge invariant
form as
\begin{eqnarray}
\tilde{\varepsilon_i}=\varepsilon_i
 -\frac{(k_1\!\cdot\! \varepsilon_i)}{(k_1\!\cdot\! p_i)} p_i.
  \label{fac6}
\end{eqnarray}
The introduction of the gauge invariant wave vector considerably
simplifies the amplitude form and direcltly justifies gauge
invariance. The second bracket term of Eq.\ (\ref{fac5}) can be
shown to be the same as the Compton scattering amplitude of a
charged vector boson given by
\begin{eqnarray}
{\cal M}_{\gamma W}&=&-2e^2
      \left[(\tilde{\epsilon_1}\!\cdot\!\tilde{\epsilon^*_2})
      (\tilde{\varepsilon_1}\!\cdot\!\tilde{\varepsilon^*_2})
      \!-(k_1\!\cdot\! k_2)\!
      \left\{\frac{(\tilde{\epsilon_1}
      \!\cdot\!\tilde{\varepsilon_1})
      (\tilde{\epsilon^*_2}\!\cdot\!
      \tilde{\varepsilon^*_2})}{p_1\!\cdot\! k_1}
   -\frac{(\tilde{\epsilon_1}\!\cdot\!\tilde{\varepsilon^*_2})
      (\tilde{\epsilon^*_2}\!\cdot\!
      \tilde{\varepsilon_1})}{p_1\!\cdot\! k_2}
      \right\}\right.\nonumber\\
    &&\left. \hskip 1.9cm
      -\frac{1}{p_1\!\cdot\! k_2}
      \{(\tilde{\epsilon^*_2}\!\cdot\!\tilde{\varepsilon_1})
      (p_2\!\cdot\!\tilde{\epsilon_1})
      (k_2\!\cdot\!\tilde{\varepsilon^*_2})
      +(\tilde{\epsilon_1}\!\cdot\!\tilde{\varepsilon^*_2})
      (p_2\!\cdot\!\tilde{\epsilon^*_2})
      (k_2\!\cdot\!\tilde{\varepsilon_1})
     \right.\nonumber\\
    &&\left. \hskip 1.9cm
      -(\tilde{\epsilon_2}\!\cdot\!\tilde{\varepsilon^*_2})
      (p_2\!\cdot\!\tilde{\epsilon_1})
      (k_2\!\cdot\!\tilde{\varepsilon_1})\}
      +\frac{1}{p_1\!\cdot\! k_1}
      (\tilde{\epsilon_1}\!\cdot\!\tilde{\varepsilon_1})
      (p_2\!\cdot\!\tilde{\epsilon^*_2})
      (k_2\!\cdot\!\tilde{\varepsilon^*_2})\right].
\label{Mrv}
\end{eqnarray}

Consequently, the amplitude ${\cal M}_{gW\rightarrow \gamma W}$
is expressed in a factorized form as
\begin{eqnarray}
{\cal M}_{gW\rightarrow \gamma W}
  &=&\frac{\kappa}{2e} F
   \left(\frac{p_2\cdot\tilde{\epsilon_1}}{p_2\cdot k_1}\right)
   \left[{\cal M}_{\gamma W}\right].
\end{eqnarray}
\subsection{Gravitational Compton scattering}

In this subsection we investigate the factorization property
of the gravitational elastic processes $gX\rightarrow gX$ for
$X=s$, $f$, or $W$.
The Feynman diagrams for these processes are shown in Fig.~3,
where the solid line is for $X$.
Through this subsection the $k_1(k_2)$ denote 4-momenta of the
incident(final) graviton and $p_1(p_2)$ denote 4-momenta of the
incident(final) $X$ particle.

\subsubsection{\mbox{$gs\rightarrow gs$}}

The relevant Lagrangian of the process $gs\rightarrow gs$ consists
of five parts;
\begin{eqnarray}
{\cal L}^{gs}_{I}= {\cal L}_g^0
      +\kappa {\cal L}_g^1+ {\cal L}_{gs}^0
      +\kappa {\cal L}_{gs}^1+\kappa^2{\cal L}_{gs}^2 .
\end{eqnarray}
The transition amplitude after absorbing the contact term
denoted by the last diagram in Fig.~3 becomes
\begin{eqnarray}
{\cal M}_{gs}={\cal M}^{gs}_a+{\cal M}^{gs}_b
             +{\cal M}^{gs}_c,
\end{eqnarray}
\begin{eqnarray}
&&{\cal M}^{gs}_a=-\frac{\kappa^2}{4} \frac{1}{(q_1^2-m^2)}
       \Bigl[ p_1\cdot k_1 (\epsilon_1\cdot\epsilon^*_2)
      -2(p_2\cdot\epsilon^*_2)( p_1\cdot\epsilon_1)\Bigr]^2,\\
&&{\cal M}^{gs}_b=-\frac{\kappa^2}{4} \frac{1}{(q_2^2-m^2)}
       \Bigl[ p_2\cdot k_1 (\epsilon_1\cdot\epsilon^*_2)
      +2(p_2\cdot\epsilon_1)( p_1\cdot\epsilon^*_2)\Bigr]^2,\\
&&{\cal M}^{gs}_c=-\frac{\kappa^2}{4} \frac{1}{q_3^2}
       \Bigl[[(p_2+p_1)\cdot k_1](\epsilon_1\cdot\epsilon^*_2)
      +2(p_2\cdot\epsilon_1)(p_1\cdot\epsilon^*_2)
      -2(p_2\cdot\epsilon^*_2)(p_1\cdot\epsilon_1)\Bigr]^2.
\end{eqnarray}
It is now straightforward to determine the corresponding factors
$A^{gs}_i$ and $B^{gs}_i$ with the same $C^{gs}_i$ $(i=1,2,3)$
as in Eq.~(\ref{Ci});
\begin{eqnarray}
A^{gs}_i&=&\kappa^2 B^{\gamma s}_i,\ \
B^{gs}_i=-\frac{B^{\gamma s}_i}{4}\ \ (i=1,2,3).
\end{eqnarray}

Consequently we obtain the factorized transition amplitude as
\begin{eqnarray}
{\cal M}_{gs}=\frac{\kappa^2}{8e^4}F[{\cal M}_{\gamma s}]^2.
             \label{fac22}
\end{eqnarray}
Note that the resulting amplitude is exactly the square of the
standard scalar-Compoton scattering amplitude.

\subsubsection{\mbox{$gf\rightarrow gf$}}

The relevant Lagrangian for the process $gf\rightarrow gf$ is
made up of five parts;
\begin{eqnarray}
{\cal L}^{gf}_{I}={\cal L}_g^0+\kappa {\cal L}_g^1
         + {\cal L}_{gf}^0
         +\kappa {\cal L}_{gf}^1+ \kappa^2 {\cal L}_{gf}^2.
\end{eqnarray}
The transition amplitude ${\cal M}_{gf}$ for the process
in Fig.~3 is reorganized after absorbing the contact term
into other three parts as
\begin{eqnarray}
{\cal M}_{gf}={\cal M}^{gf}_a+{\cal M}^{gf}_b+{\cal M}^{gf}_c,
\end{eqnarray}
\begin{eqnarray}
&&{\cal M}^{gf}_a=\frac{\kappa^2}{8(q^2_1-m^2)}
  \left[2(\epsilon_1\cdot p_1)(\epsilon^*_2\cdot p_2)
  -(\epsilon_1\cdot\epsilon^*_2)(p_1\cdot k_1)\right]\nonumber\\
&&\hskip 2.7cm \times \left[\bar{u}(p_2,s_2)
  \left[\not\!{\epsilon^*_2}
   (\not\!{q_1}+m)\not\!{\epsilon_1}\right]u(p_1,s_1)\right],\\
&&{\cal M}^{gf}_b=\frac{\kappa^2}{8(q^2_2-m^2)}
   \left[2(\epsilon_1\cdot p_2)
   (\epsilon^*_2\cdot p_1)+(\epsilon_1\cdot \epsilon^*_2)
   (p_1\cdot k_2)\right]\nonumber\\
  &&\hskip 2.7cm \times\left[\bar{u}(p_2,s_2)\left[\not\!{\epsilon_1}
   (\not\!{q_2}+m)\not\!{\epsilon_2}
    \right]u(p_1,s_1)\right],\nonumber\\
&&{\cal M}^{gf}_c=-\frac{\kappa^2}{4q_3^2}\left[2
    (\epsilon_1\cdot p_2)(\epsilon^*_2\cdot p_1)
    -2(\epsilon_1\cdot p_1)(\epsilon^*_2\cdot p_2)
    +(\epsilon_1\cdot\epsilon^*_2)(p_1+p_2)\cdot k_1)\right]
     \nonumber\\
     &&\hskip 1.9cm\times
    \left[\bar{u}(p_2,s_2)
    \left[\not\!{\epsilon_1}(\epsilon^*_2\cdot k_1)
    +\not\!{\epsilon^*_2}(\epsilon_1\cdot k_2)
    -(\epsilon_1\cdot\epsilon^*_2) \not\!{k_2}\right]
    u(p_1,s_1)\right].
\end{eqnarray}
We obtain the factors $A^{gf}_i$ and $B^{gf}_i$ ($i=1,2,3$) as
\begin{eqnarray}
A^{gf}_i=\kappa^2 B^{\gamma s}_i,\ \
B^{gf}_i=\frac{B^{\gamma f}}{4}\ \ (i=1,2,3).
\end{eqnarray}

As a result the transition amplitude is reduced to the factorized
form;
\begin{eqnarray}
{\cal M}_{gf}&=&\frac{\kappa^2}{8e^4} F
                \left[{\cal M}_{\gamma s}\right]
                \left[{\cal M}_{\gamma f}\right],
\end{eqnarray}
where the expressions of ${\cal M}_{\gamma s}$ and
${\cal M}_{\gamma f}$ are given in Eq.\ (\ref{fac4}) and
Eq.\ (\ref{Mrf}), respectively.
Note that the transition amplitude of graviton-fermion
scattering is factorized into the transition amplitude of
scalar-Compton scattering amplitude and the fermion-Compton
scattering amplitude.

\subsubsection{\mbox{$gW\rightarrow gW$}}

The relevant Lagrangian for the process $gW\rightarrow gW$ is
composed of five parts up to O($\kappa^2$);
\begin{eqnarray}
{\cal L}^{gW}_{I}={\cal L}_g^0+\kappa {\cal L}_g^1
            + {\cal L}_{gW}^0
            +\kappa {\cal L}_{gW}^1+ \kappa^2 {\cal L}_{gW}^2.
\end{eqnarray}
The amplitude expression of the process is so complicated that
the explicit presentation will be omitted.
Instead, we write down just the factors $A^{gW}_i$ and $B^{gW}_i$
($i=1,2,3$) given by the realtions
\begin{eqnarray}
A^{gW}_i=-\kappa^2 B_i^{\gamma s},
\ \ B^{gW}_i=\frac{B^{\gamma W}_i}{4}.
\end{eqnarray}
It is now clear that the amplitude of graviton scattering
with a vector boson can be written in a factorized form as
\begin{eqnarray}
{\cal M}_{gW}&=&\frac{\kappa^2}{8e^4}
               F\left[{\cal M}_{\gamma s}\right]
                \left[{\cal M}_{\gamma W}\right],
\end{eqnarray}
where ${\cal M}_{\gamma s}$ and ${\cal M}_{\gamma W}$ are the
same as Eq.\ (\ref{fac4}) and Eq.\ (\ref{Mrv}), respectively.

\subsection{Graviton-graviton elastic scattering}

The process $gg\rightarrow gg$ is a pure gravitational process
of order of $\kappa^2$. The relevant Lagrangian for the process
$gg\rightarrow gg$ is made up of three terms as
\begin{eqnarray}
{\cal L}^{gg}_{I}={\cal L}_g^0+\kappa {\cal L}_g^1
   +\kappa^2 {\cal L}_g^2.
\end{eqnarray}
As in other cases, after absorbing the contact term denoted by
the last diagram in Fig.~4, we obtain the factors $A^{gg}_i$,
$B^{gg}_i$ and $C^{gg}_i$ for the process $gg\rightarrow gg$
as follows
\begin{eqnarray}
A^{gW}_i=-\kappa^2B_i,\ \ B^{gW}_i=\frac{B_i}{16}\ \
(i=1,2,3).
\end{eqnarray}
Likewise, the transition amplitude of graviton-graviton scattering
\cite{Sannan} is factorized as
\begin{eqnarray}
{\cal M}_{gg}=\frac{\kappa^2}{8e^4}
              F\left[{\cal M}_{\gamma v}\right]^2.
  \label{fac8}
\end{eqnarray}
Here, $p_1(p_2)$ and $\varepsilon_1^\mu(\varepsilon_2^\mu)$ are
the four-momentum and wave vector of another initial(final)
graviton, respectively.

\subsection{Summary}

In the present subsection, we summarize our results for the
transition amplitudes obtained from the factorization procedure;
\begin{itemize}
\item[(1)] The transition amplitudes
      ${\cal M}_{gs\rightarrow\gamma s},
      {\cal M}_{gf \rightarrow \gamma f}$ and
      ${\cal M}_{gW\rightarrow\gamma W}$ have a common factor
   \begin{eqnarray}
   \left(\frac{p_2\cdot\tilde{\epsilon}_1}{p_2\cdot k_1}\right).
   \end{eqnarray}
\item[(2)] The transition amplitudes ${\cal M}_{gs\rightarrow gs},
      {\cal M}_{gf \rightarrow gf}$, and
      ${\cal M}_{gW\rightarrow gW}$
      have as a common factor
     \begin{eqnarray}
     {\cal M}_{\gamma s}=2e^2
      \left(\tilde{\epsilon}_1\cdot\tilde{\epsilon}_2^*\right).
     \end{eqnarray}
\item[(3)] On the other hand, the graviton-graviton scattering
      amplitude ${\cal M}_{gg\rightarrow gg}$ is proportional
      to the square of the amplitude ${\cal M}_{\gamma v}$.
\item[(4)] All the transition amplitudes have as a common
     kinematical factor
     \begin{eqnarray}
     F=\frac{(p_1\cdot k_1)(p_1\cdot k_2)}{(k_1\cdot k_2)}.
     \end{eqnarray}
\item[(5)] The other factors are exactly of the same form as the
     amplitudes  ${\cal M}_{\gamma s}$, ${\cal M}_{\gamma f}$ and
     ${\cal M}_{\gamma W}$  except for their overall coupling
     constants according to the number of involved gravitons.
\item[(6)] While the form of Eq.~(\ref{fac4}) is independent of
     the choice of $\tilde{\epsilon}_i$, the form of the
     photon-vector boson
     scattering amplitude ${\cal M}_{\gamma v}$ can be modified
     if the $\tilde{\epsilon}_i$ and $\tilde{\varepsilon}_i$ are
     defined in a different way. Nevetheless, we find that the
     transiton amplitude ${\cal M}_{gg}$ satisfies Bose and
     crossing symmetries as expected from the cyclic property
     of the factorization.
     The completely symmetric expression of the amplitude
     ${\cal M}_{gg}$  can be found in Ref.~\cite{Sannan}.
\end{itemize}
%


\section{Polarization}
\label{sec:polarization}

Clearly factorization will allow us to describe every 4-body
graviton interaction with a scalar, a fermion, a photon,
a vector boson, or a graviton itself through well-known 4-body
photon interaction processes in the ordinary QED.
One noteworthy advantage from factorization is a simple explanation
for polarization phenomena in the graviton processes.

A natural cartesian basis for a polarization vector
$\epsilon^\mu(\lambda)$ with a momentum $k_1$ can be given in terms
of two arbitrary four momenta $p_1$ and $p_2$. For simplicity,
we use $k_2$, $p_1$, and $p_2$ satisfying the constraints
$k_2^2=0$ and $p_1^2=p_2^2=m^2$.
Then we can choose the basis consisting of two orthonomal
four-vectors $n_1$ and $n_2$ such that \cite{Calkul,Ginzburg}
\begin{eqnarray}
n_1^\mu=\frac{N}{2} \left[(p_1+p_2)^\mu
       -\frac{p_1\cdot (k_1+k_2)}{k_1\cdot k_2}(k_1+k_2)^\mu
       \right],\ \
n_2^\mu=N\frac{\epsilon^{\mu\nu\alpha\beta}
       {p_1}_\nu {k_2}_\alpha {k_1}_\beta}{k_1\cdot k_2},
\label{basis}
\end{eqnarray}
with the conditions
\begin{eqnarray}
k_i\cdot n_j=0,\ \ p_i\cdot n_2=0,\ \ n_i\cdot n_j=-\delta_{ij}
\ \ (i,j=1,2),
\label{435}
\end{eqnarray}
and the normalization factor $N=1/\sqrt{2F-m^2}$.
In the basis we can introduce as a polarization vector
\begin{eqnarray}
\epsilon^\mu(\lambda)=\frac{1}{\sqrt{2}}
                      \left[n_1^\mu+i\lambda n_2\right],
\end{eqnarray}
where $\lambda=\pm 1$ is for the right-and left-handed
polarization, respectively. Note that the scalar product
$(n_1\cdot \epsilon )$ of the polarization vector
$\epsilon^\mu(\lambda)$ and the four-vector $n_1$ is independent
of the helicity value $\lambda$. Certainly one can take another
set of $(n_1^\prime, n_2^\prime)$ as a basis, but it is different
from the set $(n_1, n_2)$ simply by a complex phase, which can be
neglected without any change in physical observables.

Factorization allows us to use the well-known polarization
effects in the ordinary QED for the investigation of polarization
effects in the graviton processes.  As a preliminary part, we
consider the processe  $\gamma s \rightarrow Ys$ where $Y$ is
a massless scalar. The transition amplitude of the process
$\gamma s \rightarrow Ys$ is
\begin{eqnarray}
{\cal M} (\gamma_\lambda s\rightarrow Ys)
&=&e^2\epsilon^\mu(\lambda)\left[\frac{p_1}{k_1\cdot p_1}
              -\frac{p_2}{k_1\cdot p_2}\right]_\mu
=-\frac{1}{\sqrt{2}FN},
\end{eqnarray}
where the coupling constant of the $Yss$ vertex is assumed to be $e$.
Note that the amplitude is completely independent of photon
helicity $\lambda$.

One can now show that the process $gX\rightarrow\gamma X$ with
${\cal M} (\gamma s \rightarrow Ys)$ as a factor exhibits the
same polarization property as $\gamma X\rightarrow\gamma X$
with $X$ a scalar, a fermion or a vector boson. However, the
kinematical factor in the center of mass frame takes the form;
\begin{eqnarray}
F\cdot\frac{1}{\sqrt{2}FN}=\sqrt{F-\frac{m^2}{2}}
        =\sqrt{\frac{s}{2}}\cot\frac{\theta}{2},
 \ \ \theta=\angle (g,\gamma),
\label{sqrts}
\end{eqnarray}
so that it makes the angular distribution of the graviton
process different from the corresponding QED process.
The former is more forwardly peaked than the latter.
In addition, the graviton cross section increases and might
violate unitarity at very high energy due to the $\sqrt{s}$
factor in Eq.\ (\ref{sqrts}).

Let us now consider the elastic photon-scalar scattering process
$\gamma s\rightarrow \gamma s$.
The transition amplitude is given by
\begin{eqnarray}
{\cal M}(\gamma s \rightarrow \gamma s)
&=&2e^2
    \left[(\epsilon_1\cdot\epsilon^*_2)
 -\frac{(p_1\cdot\epsilon_1)(p_2\cdot\epsilon^*_2)}{p_1\cdot k_1}
 +\frac{(p_2\cdot\epsilon_1)(p_1\cdot\epsilon^*_2)}{p_2\cdot k_1}
\right].
\end{eqnarray}
While in general two different helicity bases are needed for
two photons only one helicity basis can be employed in the
present case;
\begin{eqnarray}
\epsilon_1^\mu(\lambda)=\epsilon_2^{*\mu}(\lambda)
                =\frac{1}{\sqrt{2}}(n_1^\mu+i\lambda n_2).
\end{eqnarray}
These enable us to derive the result\cite{Cho}
\begin{eqnarray}
{\cal M}\left(\gamma_\lambda s\rightarrow\gamma_{\lambda^\prime}s
    \right)
  =2e^2\left(\delta_{\lambda\lambda^\prime}-\frac{m^2}{2F}\right).
\label{rsrs}
\end{eqnarray}
Despite scattering the photon helicity is preserved in the
massless case.

Combined with  the factorization property in the previous section,
Eq.\ (\ref{rsrs}) leads to
\begin{eqnarray}
{\cal M}(g_\lambda X\rightarrow g_{\lambda^\prime} X)
\propto F\left(\delta_{\lambda \lambda^\prime}
      -\frac{m^2}{2F}\right)
\cdot\Bigl[{\cal M}(\gamma_\lambda X\rightarrow
     \gamma_{\lambda^\prime} X)\Bigr],
\end{eqnarray}
with $X$ a scalar, a fermion, or a vector boson.
The result yields an interesting fact that in the massless case
the final graviton helicity should be the same as the initial
graviton helicity irrespective of the spin configuration of
matter fields.

On the other hand, the transition amplitude for the process
$gg\rightarrow gg$ has neither ${\cal M}(\gamma s\rightarrow Y s)$
nor ${\cal M}(\gamma s\rightarrow\gamma s)$.
As a result graviton helicity in the process $gg\rightarrow gg$
might be not preserved unlike in the processes
$gX\rightarrow\gamma X$ and $gX\rightarrow gX$.

 From now on we investigate in more detail polarization effects in
the graviton scattering processes.
As shown above, the helicity formalism permits a simple and
general understanding of the polarization effects in the graviton
scattering processes. However, it is often convenient to employ
the so-called covariant polarization density matrix formalism,
especially for a mixed state.
In the massive case, the helicity formalism requires fixing the
reference frame and has more complicated crossing symmetries.
These problems can be avoided by the covariant density matrix
formalism. In the light of these advantageous aspects the covariant
density matrix formalism is employed in the present work to get
general information on polarization effects in arbitrary reference
frame.

The polarization of a photon (or a massless spin-1 particle) beam
is completely described\cite{Ginzburg,Choi11}
in terms of the so-called Stokes parameters (STP) $\xi^\gamma_i$
($i=1,2,3$).  In the helicity basis, $\xi^\gamma_2$ is the degree
of circular polarization and the others are degrees of linear
polarization. Because a graviton has only two helicity values,
one can introduce the so-called graviton  STP
$\xi^g_i$ ($i=1,2,3$)\cite{Choi22}. Similarily, $\xi^g_2$ is for
the degree of graviton circular polarization and the others are for
degrees of graviton linear polarization. On the whole, the photon
or graviton polarization density matrix $\rho_{V}$ $(V=\gamma,g)$
is given in the helicity basis by
\begin{eqnarray}
\rho_{V}=\frac{1}{2}\left[\begin{array}{cc}
              1+\xi^{V}_2 & -\xi^{V}_3+i\xi^{V}_1 \\
               -\xi^{V}_3-i\xi^{V}_1 & 1-\xi^{V}_2
               \end{array}\right],
\end{eqnarray}
and, for a given polarization density matrix, the STP are
determined by the following relations
\begin{eqnarray}
\xi_1^{V}=-{\rm Tr}(\sigma_2\rho_V),\ \
\xi_2^{V}={\rm Tr}(\sigma_3\rho_V),\ \
\xi_3^{V}=-{\rm Tr}(\sigma_1\rho_V),\label{pol2}
\end{eqnarray}
where $\sigma_i$ ($i=1,2,3$) are the Pauli matrices.

In the covariant density matrix formalism the photon projection
operator $\epsilon^\mu(\lambda)\epsilon^{*\nu}(\lambda^\prime)$
for an incident photon beam is replaced by its photon covariant
density matrix
\begin{eqnarray}
\rho_\gamma^{\mu\nu}&=&\frac{1}{2}
  \Bigl[(n_1^\mu n_1^\nu+n_2^\mu n_2^\nu)
       -(n_1^\mu n_2^\nu+n_2^\mu n_1^\nu)\xi^\gamma_1\nonumber\\
     &&+i(n_2^\mu n_1^\nu-n_1^\mu n_2^\nu)\xi^\gamma_2
       +(n_2^\mu n_2^\nu-n_1^\nu n_1^\mu)\xi^\gamma_3\Bigr].
\label{447}
\end{eqnarray}
In the graviton case the covariant density matrix
$\rho_g^{\mu\alpha\, :\, \nu\beta}$, which should substitute
for the graviton projection operator
$\epsilon^{\mu\alpha}(\lambda)\epsilon^{*\nu\beta}(\lambda^\prime)$,
is written in terms of the graviton STP $\xi^g_i$
($i=1,2,3$)\cite{Choi22} as follows
\begin{eqnarray}
&&\rho^{\mu\alpha :\nu\beta}
       =\sum_{\lambda\lambda^\prime}\epsilon^{\mu\alpha}
        (p,\lambda)\rho_{\lambda\lambda^\prime}
           \epsilon^{*\nu\beta}(p,\lambda^\prime)\nonumber\\
&& =\frac{1}{4}
    \Bigl\{(n^\mu_1 n^\nu_1+n^\mu_2 n^\nu_2)
         (n^\alpha_1 n^\beta_1+n^\alpha_2 n^\beta_2)
        -(n^\mu_2 n^\nu_1-n^\mu_1 n^\nu_2)
         (n^\alpha_2 n^\beta_1-n^\alpha_1 n^\beta_2)\nonumber\\
&&  \hskip 0.7cm -
    \left[(n^\mu_2 n^\nu_2-n^\mu_1 n^\nu_1)
          (n^\alpha_2 n^\beta_2-n^\alpha_1 n^\beta_1)
         -(n^\mu_1 n^\nu_2+n^\mu_2 n^\nu_1)
          (n^\alpha_1 n^\beta_2+n^\alpha_2 n^\beta_1)\right]
           \xi^g_3\nonumber\\
&&  \hskip 0.6cm
    +i\left[(n^\mu_1 n^\nu_1+n^\mu_2 n^\nu_2)
            (n^\alpha_2 n^\beta_1-n^\alpha_1 n^\beta_2)
           +(n^\mu_2 n^\nu_1-n^\mu_1 n^\nu_2)
            (n^\alpha_1 n^\beta_1+n^\alpha_2 n^\beta_2)\right]
             \xi^g_2 \nonumber\\
&&   \hskip 0.7cm -
    \left[(n^\mu_2 n^\nu_2-n^\mu_1 n^\nu_1)
          (n^\alpha_1 n^\beta_2+n^\alpha_2 n^\beta_1)
         -(n^\mu_1 n^\nu_2+n^\mu_2 n^\nu_1)
          (n^\alpha_2 n^\beta_2-n^\alpha_1 n^\beta_1)\right]
           \xi^g_1\Bigr\}.
   \label{gcd}
\end{eqnarray}

In a process with an incident graviton the transition amplitude
can be written as
\begin{eqnarray}
{\cal M_I}=\epsilon_{\mu\nu}{\cal A_I^{\mu\nu}},
\end{eqnarray}
and then the absolute square of the amplitude is given by
\begin{eqnarray}
|{\cal M_I}|^2=\epsilon^{\mu\alpha}\epsilon^{*\nu\beta}
     {\cal A_I}_{\mu\alpha}{\cal A_I^*}_{\nu\beta}.\label{square}
\end{eqnarray}
Polarization effects of an incident graviton beam are determined
by replacing $\epsilon^{\mu\alpha}\epsilon^{*\nu\beta}$ in
Eq.~(\ref{square}) by the covariant density matrix
$\rho^{\mu\alpha:\nu\beta}$ in Eq.\ (\ref{gcd}).

On the other hand, the scattering amplitude for a graviton
production is in general written as
\begin{eqnarray}
{\cal M_F}(\lambda)=\epsilon^*_{\mu\nu} (p,\lambda)
         {\cal A_F^{\mu\nu}}.
\end{eqnarray}
Then the final spin-2 polarization density matrix
$\rho_{\lambda\lambda^\prime}$ is determined through the relation
\begin{eqnarray}
\rho_{\lambda\lambda^\prime}
       =\frac{\cal{M}_F(\lambda)\cal{M}^*_F(\lambda^\prime)}
        {\sum_{\lambda\lambda^\prime}
        \cal{M}_F(\lambda)\cal{M}^*_F(\lambda^\prime)}.
 \label{PolD}
\end{eqnarray}
After such manipulation, Eq.\ (\ref{pol2}) is used to obtain
the final graviton STP.

In the following we present the differential
cross sections in a $2\times 2$ matrix form in order to
consider the beam interference effects and to relate directly
those expressions with the final polarization density matrices
through the relation (\ref{PolD}).

\subsection{Graviton conversion into a photon}

In this subsection, we use the factorized amplitudes for
the processes $g X \rightarrow \gamma X$ obtained in Sec. III A
in order to consider the polarization effects.
Since those processes have the amplitude
${\cal M}_{\gamma s\rightarrow Ys}$ as a common factor,
it is obvious that  the polarization effects of the initial
graviton beam should be identical to those of the initial
photon beam in the process $\gamma X\rightarrow \gamma X$.

\subsubsection{\mbox{$gs\rightarrow \gamma s$}}

First of all, we consider the simplest process
$gs\rightarrow\gamma s$. Following the procedure described before,
we obtain the differential cross section of the process
$gs\rightarrow \gamma s$ in a $2\times 2$ matrix form as
\begin{eqnarray}
\frac{d\sigma^{\gamma s}}{dt}(\lambda\lambda^\prime)
     =\frac{\pi\alpha\alpha_g}{4}\frac{(su-m^4)}{(s-m^2)^2 t}
       \left(F^{\gamma s}_0
     +\sum^{3}_{i=1}
      F^{\gamma s}_i\chi_i\right)_{\lambda\lambda^\prime},
\end{eqnarray}
\begin{eqnarray}
&&F^{\gamma s}_0=1+2f+2f^2 +2f(1+f)\xi^g_3, \ \
  F^{\gamma s}_1=(1+2f)\xi^g_1, \nonumber\\
&&F^{\gamma s}_2=(1+2f)\xi^g_2, \ \
  F^{\gamma s}_3=(1+2f+2f^2)\xi^g_3+2f(1+f),
\label{Fgrs}
\end{eqnarray}
where we have introduced the notations
\begin{eqnarray}
&&\alpha=\frac{e^2}{4\pi},\ \ \alpha_g=\frac{\kappa^2}{4\pi},\ \
  f=-\frac{m^2}{2F},\nonumber\\
&&s=(p_1+k_1)^2,\ \ u=(p_1-k_2)^2,\ \ t=(p_1-p_2)^2.
\end{eqnarray}
Here the $\chi_i$ ($i=1,2,3$) are three $2\times 2$ matrices
related with the Pauli matrices $\sigma_i$ as
\begin{eqnarray}
\chi_1=-\sigma_2, \  \ \chi_2=\sigma_3, \  \ \chi_3=-\sigma_1.
\end{eqnarray}
Then the final photon STP can be obtained from Eq.~(\ref{Fgrs})
as
\begin{eqnarray}
\xi_i^{\prime\gamma}=\frac{F^{\gamma s}_i}{F^{\gamma s}_0}.
\label{XI}
\end{eqnarray}
The polarization of the final photon depends on that of the
incident graviton in general.
One can now check that the final photon STP are identical to
those of the initial graviton beam in the massless case,
i.e. $\xi_i^{\prime\gamma}=\xi^g_i$ ($i=1,2,3$).
This result is due to the fact that the amplitude has not only
the factor ${\cal M}_{\gamma s\rightarrow Ys}$ but also the
factor ${\cal M}_{\gamma s\rightarrow\gamma s}$.

\subsubsection{\mbox{$gf\rightarrow \gamma f$}}

In the process $gf\rightarrow\gamma f$ we can in principle
consider the case where all the particles are polarized.
But in order to look into the implications
from factorization to the polarizations, it will be sufficient
to consider the case where all the other particles except the final
fermion are polarized.

For notational convenience we first introduce the invariants
\begin{eqnarray}
a_1=\frac{(s_1\cdot k_1)}{m},\ \ a_2=\frac{(s_1\cdot k_2)}{m},\ \
\epsilon=\frac{\varepsilon_{\mu\nu\rho\sigma}s^\mu_1 k^\nu_1
                 p^\rho_1 k^\sigma_2}{m(s-m^2)},
\end{eqnarray}
where $s_1$ is the incident fermion spin 4-vector and $m$ is the
fermion mass. $k_i$ and $p_i$ ($i=1,2$) are defined in the same way
as in Section.~III.

The differential cross section of the process
$gf\rightarrow\gamma f$ is obtained in a $2\times 2$ matrix
form as
\begin{eqnarray}
\frac{d\sigma^{\gamma f}}{dt}(\lambda\lambda^\prime)
     =\frac{\pi\alpha\alpha_g}{4}\frac{(su-m^4)}{(s-m^2)^2 t}
      \left(F^{\gamma f}_0
     +\sum^{3}_{i=1}
      F^{\gamma f}_i\chi^i\right)_{\lambda\lambda^\prime},
\label{gfrfd}
\end{eqnarray}
\begin{eqnarray}
&&F^{\gamma f}_0=-h_1+4f(1+f)(1-\xi_3^g)
    -2f\left[(1+2f)a_1+a_2\right]\xi_2^g, \nonumber\\
&&F^{\gamma f}_1=2(1+2f)\xi_1^g
    -4fh_2\epsilon\xi_2^g,\nonumber\\
&&F^{\gamma f}_2=-(1+2f)\left[h_1\xi_2^g
    +2(1-\xi^g_3)fa_1\right]
    -2\frac{fa_2}{h_2}\xi_3^g-4f\epsilon\xi_1^g,\nonumber\\
&&F^{\gamma f}_3=-4f(1+f)+2\left[1+2f(1+f)\right]\xi_3^g
    +2f\left[(1+2f)a_1+h_2 a_2\right]\xi_2^g,
\label{Fgf}
\end{eqnarray}
where in addition to $f$ we introduce two Lorentz invariant
functions
\begin{eqnarray}
h_1=\frac{s-m^2}{u-m^2}+\frac{u-m^2}{s-m^2},\ \
h_2=\frac{s-m^2}{u-m^2}.
\end{eqnarray}

When averaged over the initial spin states, Eq.~(\ref{gfrfd})
gives the same results as in Ref.~\cite{Voronov}.
 From the ratio of $F^{\gamma f}_i$ to $F^{\gamma f}_0$ in
Eq.~(\ref{Fgf}) similar to Eq.~(\ref{XI}), we can obtain the STP
of the final photon beam.
We note that when the initial graviton STP
are used in place of the photon STP, the final photon polarization
is identical to that\cite{Ginzburg} of the QED Compton scattering
process.  This is a result expected from the factorization property.
For the case of massless and unpolarized fermion, we obtain the
photon STP as
\begin{eqnarray}
\xi_1^{\prime\gamma}=-\left(\frac{2su}{s^2+u^2}\right)\xi^g_1,\ \
\xi_2^{\prime\gamma}= \xi^g_2,\ \
\xi_3^{\prime\gamma}=-\left(\frac{2su}{s^2+u^2}\right)\xi^g_3.\ \
\end{eqnarray}
As a result, the degree of circular polarization is preserved
despite scattering even in the graviton-fermion scattering process.
However, the degrees of linear polarization are reduced according to
the scattering angle.

\subsubsection{\mbox{$gW\rightarrow \gamma W$}}

In order to describe the process with all polarized particles,
we have to introduce four different sets of Stokes parameters.
For simplicity, we consider the case where the initial and final
massive vector bosons are unpolarized.
Then in a $2\times 2$ matrix form, the differential cross section
of the process $gW\rightarrow\gamma W$ is given by
\begin{eqnarray}
\frac{d\sigma^{\gamma W}}{dt}(\lambda\lambda^\prime)
   =\frac{\pi\alpha\alpha_g}{12}\frac{(su-m^4)}{(s-m^2)^2 t}
    \left(F^{\gamma W}_0
   +\sum^{3}_{i=1}
    F^{\gamma W}_i\chi^i\right)_{\lambda\lambda^\prime},
\end{eqnarray}
\begin{eqnarray}
&&F^{\gamma W}_0=6(1+\xi_3^g)f(1+f)+3+4h_1+2h^2_1,\nonumber\\
&&F^{\gamma W}_1=-3(1+2f)\xi_1^g,\ \
  F^{\gamma W}_2=(1+2f)(5+4h_1)\xi_2^g, \nonumber\\
&&F^{\gamma W}_3=3(1+2f+2f^2)\xi_3^g+6f(1+f).
\end{eqnarray}

In the massless case where $f$=0, the final photon STP
are shown to be
\begin{eqnarray}
&&\xi^{\prime\gamma}_1=-\left(\frac{3}{2h^2_1
       +4h_1+3}\right)\xi^g_1,\ \
\xi^{\prime\gamma}_2=\left(\frac{4h_1+5}{2h^2_1
       +4h_1+3}\right)\xi^g_2,\ \
\xi^{\prime\gamma}_3=\left(\frac{3}{2h^2_1
       +4h_1+3}\right)\xi^g_3.\nonumber\\
&&{ }
\end{eqnarray}
Note that even in the massless limit the final photon helicity
is not equal to the initial graviton helicity.

\subsection{Gravitational Compton scattering}

In this subsection we consider polarization effects in elastic
graviton scattering processes $gX\rightarrow gX$, by using the
results obtained in Sec.~III B.
Since the factorization forces every amplitude to have the
amplitude ${\cal M}_{\gamma s\rightarrow \gamma s}$ as a factor,
the graviton helicity should be preserved when the particle $X$
is massless. At first, we present the results in the massive case
and then discuss the massless limit to check graviton helicity
preservation.

\subsubsection{\mbox{$gs\rightarrow gs$}}

The simplest one of elastic graviton-matter scattering processes
is the graviton-scalar scattering process $gs\rightarrow gs$.
The differential cross section is written in a $2\times 2$ matrix
form as
\begin{eqnarray}
\frac{d\sigma^{gs}}{dt}(\lambda\lambda^\prime)
         =\frac{\pi\alpha_g^2}{16}\left(\frac{u-m^2}{t}\right)^2
          \left(F^{gs}_0
         +\sum^{3}_{i=1}
          F^{gs}_i\chi_i\right)_{\lambda\lambda^\prime},
\label{gsgsdc}
\end{eqnarray}
\begin{eqnarray}
&&F^{gs}_0=\frac{1}{2}(1+2f)^2+f^2(1+f)^2(1+\xi^g_3), \nonumber\\
&&F^{gs}_1=\frac{1}{2}(1+2f)(1+f)^2\xi^g_1, \ \
  F^{gs}_2=\frac{1}{2}(1+2f)(1+f)^2\xi^g_2, \nonumber\\
&&F^{gs}_3=\frac{1}{2}(1+2f)^2\xi^g_3+f^2(1+f)^2(1+\xi^g_3).
\label{Fggs}
\end{eqnarray}
We can see the difference in the polarization of the outgoing
graviton from the polarization of the outgoing photon in the
process $gs \rightarrow \gamma s$ by comparison of Eq.~(\ref{Fgrs})
with Eq.~(\ref{Fggs}). When interacting gravitons are unpolarized,
the differential cross section of the Eq.~ (\ref{gsgsdc}) gives
the same results as in Ref.~\cite{Berends}.
In the massless case, where $f$ vanishes, we find that
$\xi^{\prime g}_i=\xi^g_i$ ($i=1,2,3$)
as in the process $gs\rightarrow \gamma s$  so that graviton
polarization is preserved in spite of scattering.

\subsubsection{\mbox{$gf\rightarrow gf$}}

The next simplest process is $gf\rightarrow gf$.
The differential cross section of a $2\times 2$ matrix form is
given by
\begin{eqnarray}
\frac{d\sigma^{gf}}{dt}(\lambda\lambda^\prime)
     =\frac{\pi\alpha_g^2}{16}\left(\frac{u-m^2}{t}\right)^2
      \left(F^{gf}_0
     +\sum^{3}_{i=1}
      F^{gf}_i\chi_i\right)_{\lambda\lambda^\prime},
\label{gfgfdc}
\end{eqnarray}
\begin{eqnarray}
&&F^{gf}_0=-f(1+f)(3h_1-2)-h_1+4f^2(1+f)^2(1+\xi^g_3)\nonumber\\
    &&\hskip 1.3cm -\left[(1+2f+2f^2)\left\{(1+2f)a_1+a_2\right\}
      +2(1+2f)\epsilon\right]f\xi^g_2 \nonumber\\
&&F^{gf}_1= 2\left[(1+f)^4-f^4\right]\xi^g_1
      +4(1+f)f^2h_2\epsilon\xi^g_2 ,\nonumber\\
&&F^{gf}_2= -\left[(1+2f+2f^2)\left\{(1+2f)a_1+a_2\right\}
      +2(1+2f)h_2\epsilon\right]f\nonumber\\
     &&\hskip 1.3cm -(1+2f)\left[f^2 h_1+(h_1-2f)(1+f)\right]
       \xi^g_2\nonumber\\
     &&\hskip 1.3cm -2(1+f)f^2\left[2\epsilon\xi^g_1
       -\left\{(1+2f)a_1+\frac{a_2}{h_2}\right\}\xi^g_3\right],
       \nonumber\\
&&F^{gf}_3=4f^2(1+f)^2(\xi^g_3-1)+2(1+2f^2+2f^2)\xi^g_3
       \nonumber\\
    &&\hskip 1.3cm +2f^2(1+f)\left\{(1+2f)a_1
       +h_2 a_2\right\}\xi^g_2,
\label{ggf}
\end{eqnarray}
where the spin states of the final fermion are summed.
When averaged over the initial spin states, Eq.~(\ref{gfgfdc})
gives the same results as in Ref.~\cite{Voronov}.
Then the STP of the outgoing graviton can be obtained from
the Eq.~(\ref{ggf})
\begin{eqnarray}
\xi_i^{\prime g}=\frac{F^{gf}_i}{F^{gf}_0}.
\end{eqnarray}
One can see the polarization of the outgoing graviton is
influenced by the polarization of the incident graviton as well as
incident fermion. In the massless case, the final graviton STP
reduce to
\begin{eqnarray}
\xi_1^{\prime g}=-\left(\frac{2su}{s^2+u^2}\right)\xi^g_1,\ \
\xi_2^{\prime g}= \xi^g_2,\ \
\xi_3^{\prime g}=-\left(\frac{2su}{s^2+u^2}\right)\xi^g_3,\ \
\end{eqnarray}
It is straightforward to show that graviton helicity is preserved
in the massless case as in the process $gf\rightarrow\gamma f$.

\subsubsection{\mbox{$gW\rightarrow gW$}}

We consider only the case where the massive vector boson $W$ is
unpolarized. The differential cross section of the process
$gW\rightarrow gW$ is written in a $2\times 2$ matrix form as
\begin{eqnarray}
\frac{d\sigma^{gW}}{dt}(\lambda\lambda^\prime)
    =\frac{\pi\alpha_g^2}{6}\left(\frac{u-m^2}{t}\right)^2
     \left(F^{gW}_0+\sum^{3}_{i=1}
     F^{gW}_i\chi_i\right)_{\lambda\lambda^\prime},
\end{eqnarray}
\begin{eqnarray}
&&F^{gW}_0=6f^4+12f^3+2f^2(h_1-3)^2
    +2f(h_1^2-2h_1-2)+(h_1^2-1)+6f^2(1+f)^2\xi^g_3, \nonumber\\
&&F^{gW}_1=-3(1+4f+6f^2+4f^3)\xi^g_1, \nonumber\\
&&F^{gW}_2=\left[2f^2(2h_1+1)(2f+3)
     -2f(h_1^2-2h_1-2)+(h_1^2-1)\right]\xi^g_2, \nonumber\\
&&F^{gW}_3=6f^2(1+f)^2+3[f^4+(1+f)^4]\xi^g_3.
\end{eqnarray}

In the massless limit the final graviton STP become
\begin{eqnarray}
\xi_1^{\prime g}=-\left(\frac{3}{h^2_1-1}\right)\xi^g_1,\ \
\xi_2^{\prime g}= \xi^g_2,\ \
\xi_3^{\prime g}=\left(\frac{3}{h^2_1-1}\right)\xi^g_3.
\label{gWgW}
\end{eqnarray}
Clearly graviton helicity is preserved in the massless case as
shown in Eq.~(\ref{gWgW}).

\subsubsection{\mbox{$g\gamma\rightarrow g\gamma$}}

The differential cross section of the process
$g\gamma\rightarrow g\gamma$ is written in a $2\times 2$ matrix
form as
\begin{eqnarray}
\frac{d\sigma^{g\gamma}}{dt}(\lambda\lambda^\prime)
   =\frac{\pi\alpha_g^2}{32s^2}
    \left(F^{g\gamma}_0
   +\sum^{3}_{i=1}
    F^{g\gamma}_i\chi_i\right)_{\lambda\lambda^\prime},
    \label{grgrdc}
\end{eqnarray}
\begin{eqnarray}
&&F^{g\gamma}_0=\frac{1}{2t^2}\left[(1+\xi^g_2\xi^\gamma_2)s^4
                +(1-\xi^g_2\xi^\gamma_2)u^4\right],\ \
  F^{g\gamma}_1=\left(\frac{su}{t}\right)^2\xi^g_1,\nonumber\\
&&F^{g\gamma}_2=\frac{1}{2t^2}\left[(\xi^g_2+\xi^\gamma_2)s^4
                +(\xi^g_2-\xi^\gamma_2)u^4\right],\ \
  F^{g\gamma}_3=\left(\frac{su}{t}\right)^2\xi^g_3,
\end{eqnarray}
where $\xi^\gamma_i$ and $\xi^g_i$ ($i=1,2,3$) are the STP's of
the incident photon beam and the incident graviton beam,
respectively, and the final photon polarization is summed.
When interacting gravitons and photons are unpolarized, the
differential cross section of the Eq.~(\ref{grgrdc}) gives
the same results as in Ref.~\cite{Berends}.
Then the explicit form of the outgoing graviton STP can be
obtained as
\begin{eqnarray}
\xi_i^{\prime g}=\frac{F^{g\gamma}_i}{F^{g\gamma}_0}.
\end{eqnarray}
Similarly, we can obtain the change of polarization of the photon
colliding with a graviton.
After taking an average over the initial photon polarization,
we obtain the final graviton STP as
\begin{eqnarray}
\xi^{\prime g}_1=\left(\frac{2s^2u^2}{s^4+u^4}\right)\xi^g_1,\ \
\xi^{\prime g}_2=\xi^g_2,\ \
\xi^{\prime g}_3=\left(\frac{2s^2u^2}{s^4+u^4}\right)\xi^g_3.
\end{eqnarray}
We also find that the graviton helicity is preserved for the
unpolarized incident and final photon beams, but in general the
final graviton STP depends on both initial photon and graviton
STP's.

\subsection{Graviton-Graviton scattering}

In the elastic graviton-graviton scattering process,
three-graviton vertices and a four-graviton vertex are involved.
Even though the vertices are so complicated as shown explicitly
in the Appendix, we obtain the very simple differential
cross section of the process $gg\rightarrow gg$ as
\begin{eqnarray}
\frac{d\sigma^{gg}}{dt}(\lambda\lambda^\prime)
   =\frac{\pi\alpha_g^2}{32 s^4 u^2 t^2}
    \left(F^{gg}_0+\sum^{3}_{i=1}
    F^{gg}_i\chi_i\right)_{\lambda\lambda^\prime},
 \label{ggdc}
\end{eqnarray}
\begin{eqnarray}
&&F^{gg}_0=\frac{1}{2}(1+\xi^{g_1}_2\xi^{g_2}_2)s^8
      +\frac{1}{2}(1-\xi^{g_1}_2\xi^{g_2}_2)(u^8+t^8)
      +(\xi_3^{g_1}\xi_3^{g_2}+\xi_1^{g_1}\xi_1^{g_2})u^4 t^4,
       \nonumber\\
&&F^{gg}_1=(\xi_1^{g_2}u^4+\xi_1^{g_1}t^4)s^4,\nonumber\\
&&F^{gg}_2=\frac{1}{2}(\xi_2^{g_2}-\xi^{g_1}_2)(u^8-t^8)
          +\frac{1}{2}(\xi_2^{g_2}+\xi^{g_1}_2)s^8, \ \
  F^{gg}_3=(\xi_3^{g_2}u^4+\xi_3^{g_1}t^4)s^4.
\end{eqnarray}
Here $\xi_i^{g1}(\xi_i^{g2})$ are the incident graviton STP with
momentum $k_1(p_1)$ and the polarization of one of two final
graviton beams is summed.
When interacting gravitons and photons are unpolarized,
the differential cross section of the Eq.~(\ref{ggdc}) gives the
same results as in Ref.~\cite{Berends}.
In contrast to all the other processes under consideration in
the present section, the graviton helicity in the process
$gg\rightarrow gg$ is not preserved.
This reflects that the transition amplitude contains neither
${\cal M}_{\gamma s\rightarrow Ys}$ nor
${\cal M}_ {\gamma s\rightarrow\gamma s}$ as a common factor,
which can lead to the graviton helicity preservation.


\section{Summary and Discussion}
\label{sec:summary}

Gravitational gauge invariance and graviton transversality
force the transition amplitudes of four-body graviton
interactions to be factorized;
\begin{eqnarray}
\left\{\begin{array}{c}
      {\cal M}_{gs \rightarrow \gamma s}\\
      {\cal M}_{gf \rightarrow \gamma f}\\
      {\cal M}_{gW \rightarrow \gamma W}
      \end{array}\right\}
&=& -\sqrt{\frac{\alpha_g}{4\alpha}} F
   \left[{\cal M}_{\gamma s\rightarrow Ys}\right]\times
\left\{\begin{array}{c}
     {\cal M}_{\gamma s}\\
     {\cal M}_{\gamma f}\\
     {\cal M}_{\gamma W}
      \end{array}\right\},\\
&& { } \nonumber\\
\left\{\begin{array}{c}
      {\cal M}_{gs}\\
      {\cal M}_{gf}\\
      {\cal M}_{gW}
      \end{array}\right\}
&=&\frac{\alpha_g}{8\alpha} F \left[{\cal M}_{\gamma s}\right]
   \times \left\{\begin{array}{r}
      {\cal M}_{\gamma s}\\
     {\cal M}_{\gamma f}\\
      {\cal M}_{\gamma W}
      \end{array}\right\},\\
&& { } \nonumber\\
{\cal M}_{gg}&=&\frac{\alpha_g}{8\alpha}F
      \left[{\cal M}_{\gamma v}\right]\times
      \left[{\cal M}_{\gamma v}\right].
\end{eqnarray}

The introduction of manifestly gauge invariant four-vectors
$\tilde{\epsilon_i}$ and $\tilde{\varepsilon_i}$ ($i=1,2$) renders
each amplitude expression simplified.
This simplification with the factorization property justifies why,
with all the very complicated three-graviton and four-graviton
vertices\cite{Berends,DeWitt}, the final form of transition
amplitudes is so simple.

The factorized transition amplitudes faciliate the investigation
of polarization effects in the four-body graviton interactions.
The transition amplitudes for the graviton interactions with a
photon or a matter field, $gX \rightarrow \gamma X$, where $X$
is a scalar, a fermion, or a vector boson, have essentially the
same transition  amplitude structure as those
involving a photon instead of the graviton, apart from a simple
overall kinematical factor.
As a result, the polarization effects involving the graviton are
identical to those for the corresponding photon if the graviton
Stokes parameters are used in place of the photon Stokes parameters.
But the kinematical factor makes the angular distribution of the
graviton process different from that of the corresponding photon
process. On the other hand, the processes $gX\rightarrow gX$
have as a common factor the elastic photon-scalar
scattering amplitude ${\cal M}_{\gamma s}$ with the scalar mass
equal to the $X$ mass in their amplitude expressions.
This leads to the conclusion that, when the particle $X$ is
massless, the graviton helicity is preserved due to the photon
helicity conservation of the process
$\gamma s\rightarrow \gamma s$ in the massless limit.
Only the mass terms cause the graviton helicity to be flipped.

The process $gg\rightarrow gg$ does not contain
${\cal M}_{\gamma s}$ as a common factor.
This point is reflected in the fact that the graviton helicity
is not preserved in the process $gg\rightarrow gg$ in spite of
the masslessness of graviton.

The validity of factorization can become more concrete through
further extensive investigation. We point out a few aspects worth
further investigating; (i) A formal proof of factorization might
be presented. There is a factorization property of the same type
in closed string theories\cite{Kawai}. From the fact that a
closed string theory reduces to a supergravity theory in the
infinite string tension limit\cite{Scherk}, one can conclude that
this is a real proof of the factorization in the linearized
gravity. However, factorization in the string theory is due to
the independence of left-moving modes and right-moving modes,
while only gravitational gauge invariance and Lorentz invariance
are imposed in the linearized gravity.
Still, the relationship between two concepts are to be established.
(ii) Factorization is expected to hold even if matter particles
have different masses. As an example, the process
$ge\rightarrow W\nu_e$ can be considered to check this point.
(iii) It will be an interesting question
whether factorization survives against any loop
effects\cite{Donoghue}.

To conclude, factorization has such a generic property in any
Lorentz-invariant gauge theory that its more intensive and
extensive investigation is expected to provide us with some
clues for the unification of gravity with other interactions.

\section*{Acknowledgments}

The work was supported in part by KOSEF (the Korea Science and
Engineering Foundation) through the SRC program and in part by
the Korean Ministry of Education. S.Y.~Choi would like to thank
MESC (the Japanese Ministry of Education, Science and Culture)
for the award of a visiting fellowship and J.S.~Shim thanks KOSEF
for the Post-Doc. fellowship. The authors are grateful to Jungil
Lee for writing a Mathematica program to construct the simplest
forms of ${\cal L}_g^1$ and ${\cal L}_g^2$. They also thank
B.~Bullock for reading the paper carefully.

\vskip 2cm

\input prepictex
\input pictex
\input postpictex

\newdimen\scale
\scale=0.7mm

\def\glhor{
\beginpicture
\setcoordinatesystem units <\the\scale,\the\scale> point at 0 0
\setplotarea x from 0 to 0, y from 0 to 0
\setplotsymbol ({.})
\setquadratic
\plot
   .0000    .0000     .4412   1.4029    1.7764   2.1823    3.1117   1.4029
  3.5470   -.1559    3.0823   -.9353    2.6177   -.1559    3.0530   1.4029
  4.3882   2.1823    5.7235   1.4029    6.1588   -.1559    5.6941   -.9353
  5.2294   -.1559    5.6647   1.4029    7.0000   2.1823    8.3353   1.4029
  8.7706   -.1559    8.3059   -.9353    7.8412   -.1559    8.2765   1.4029
  9.6118   2.1823   10.9470   1.4029   11.3823   -.1559   10.9177   -.9353
 10.4530   -.1559   10.8883   1.4029   12.2236   2.1823   13.5588   1.4029
 14.0000    .0000
/
\endpicture}

\def\glver{
\beginpicture
\setcoordinatesystem units <\the\scale,\the\scale> point at 0 0
\setplotarea x from 0 to 0, y from 0 to 0
\setplotsymbol ({.})
\setquadratic
\plot
   .0000    .0000   -1.2025    .3782   -1.8705   1.5227   -1.2025   2.6672
   .1336   3.0403     .8017   2.6420     .1336   2.2437   -1.2025   2.6168
 -1.8705   3.7613   -1.2025   4.9058     .1336   5.2790     .8017   4.8807
   .1336   4.4824   -1.2025   4.8555   -1.8705   6.0000   -1.2025   7.1445
   .1336   7.5176     .8017   7.1193     .1336   6.7210   -1.2025   7.0942
 -1.8705   8.2387   -1.2025   9.3832     .1336   9.7563     .8017   9.3580
   .1336   8.9597   -1.2025   9.3328   -1.8706  10.4773   -1.2025  11.6219
   .0000  12.0000
/
\endpicture}

\def\glupl{
\beginpicture
\setcoordinatesystem units <\the\scale,\the\scale> point at 0 0
\setplotarea x from 0 to 0, y from 0 to 0
\setplotsymbol ({.})
\setquadratic
\plot
-10.0000  10.0000  -10.6870   8.6828  -10.2899   7.1723   -8.7794   6.7753
 -7.3551   7.5778   -7.1303   8.4664   -8.0189   8.2416   -8.8214   6.8172
 -8.4243   5.3068   -6.9139   4.9097   -5.4895   5.7122   -5.2647   6.6008
 -6.1533   6.3760   -6.9558   4.9517   -6.5588   3.4412   -5.0483   3.0442
 -3.6240   3.8467   -3.3992   4.7353   -4.2878   4.5105   -5.0903   3.0861
 -4.6932   1.5757   -3.1828   1.1786   -1.7584   1.9811   -1.5336   2.8697
 -2.4222   2.6449   -3.2247   1.2206   -2.8277   -.2899   -1.3172   -.6870
   .0000    .0000
/
\endpicture}

\def\glupr{
\beginpicture
\setcoordinatesystem units <\the\scale,\the\scale> point at 0 0
\setplotarea x from 0 to 0, y from 0 to 0
\setplotsymbol ({.})
\setquadratic
\plot
   .0000    .0000    -.6870   1.3172    -.2899   2.8277    1.2206   3.2247
  2.6449   2.4222    2.8697   1.5336    1.9811   1.7584    1.1786   3.1828
  1.5757   4.6932    3.0861   5.0903    4.5105   4.2878    4.7353   3.3992
  3.8467   3.6240    3.0442   5.0483    3.4412   6.5588    4.9517   6.9558
  6.3760   6.1533    6.6008   5.2647    5.7122   5.4895    4.9097   6.9139
  5.3068   8.4243    6.8172   8.8214    8.2416   8.0189    8.4664   7.1303
  7.5778   7.3551    6.7753   8.7794    7.1723  10.2899    8.6828  10.6870
 10.0000  10.0000
/
\endpicture}

\def\gldol{
\beginpicture
\setcoordinatesystem units <\the\scale,\the\scale> point at 0 0
\setplotarea x from 0 to 0, y from 0 to 0
\setplotsymbol ({.})
\setquadratic
\plot
-10.0000 -10.0000   -8.6828 -10.6870   -7.1723 -10.2899   -6.7753  -8.7794
 -7.5778  -7.3551   -8.4664  -7.1303   -8.2416  -8.0189   -6.8172  -8.8214
 -5.3068  -8.4243   -4.9097  -6.9139   -5.7122  -5.4895   -6.6008  -5.2647
 -6.3760  -6.1533   -4.9517  -6.9558   -3.4412  -6.5588   -3.0442  -5.0483
 -3.8467  -3.6240   -4.7353  -3.3992   -4.5105  -4.2878   -3.0861  -5.0903
 -1.5757  -4.6932   -1.1786  -3.1828   -1.9811  -1.7584   -2.8697  -1.5336
 -2.6449  -2.4222   -1.2206  -3.2247     .2899  -2.8277     .6870  -1.3172
   .0000    .0000
/
\endpicture}

\def\gldor{
\beginpicture
\setcoordinatesystem units <\the\scale,\the\scale> point at 0 0
\setplotarea x from 0 to 0, y from 0 to 0
\setplotsymbol ({.})
\setquadratic
\plot
   .0000    .0000    1.3172    .6870    2.8277    .2899    3.2247  -1.2206
  2.4222  -2.6449    1.5336  -2.8697    1.7584  -1.9811    3.1828  -1.1786
  4.6932  -1.5757    5.0903  -3.0861    4.2878  -4.5105    3.3992  -4.7353
  3.6240  -3.8467    5.0483  -3.0442    6.5588  -3.4412    6.9558  -4.9517
  6.1533  -6.3760    5.2647  -6.6008    5.4895  -5.7122    6.9139  -4.9097
  8.4243  -5.3068    8.8214  -6.8172    8.0189  -8.2416    7.1303  -8.4664
  7.3551  -7.5778    8.7794  -6.7753   10.2899  -7.1723   10.6870  -8.6828
 10.0000 -10.0000
/
\endpicture}

\def\glupll{
\beginpicture
\setcoordinatesystem units <\the\scale,\the\scale> point at 0 0
\setplotarea x from 0 to 0, y from 0 to 0
\setplotsymbol ({.})
\setquadratic
\plot
-22.0000  13.0000  -20.6156  14.1867  -18.7810  14.1295  -17.8458  12.5501
-18.2392  10.7288  -19.0618  10.1880  -18.9848  11.1694  -17.5793  12.3926
-15.7447  12.3353  -14.8096  10.7559  -15.2030   8.9347  -16.0255   8.3939
-15.9486   9.3753  -14.5431  10.5984  -12.7085  10.5412  -11.7733   8.9618
-12.1667   7.1405  -12.9893   6.5997  -12.9123   7.5811  -11.5068   8.8043
 -9.6722   8.7470   -8.7371   7.1676   -9.1305   5.3464   -9.9530   4.8056
 -9.8761   5.7870   -8.4706   7.0101   -6.6360   6.9529   -5.7008   5.3735
 -6.0942   3.5522   -6.9168   3.0114   -6.8398   3.9928   -5.4343   5.2160
 -3.5997   5.1587   -2.6646   3.5793   -3.0580   1.7581   -3.8805   1.2173
 -3.8036   2.1987   -2.3981   3.4218    -.5635   3.3646     .3717   1.7852
   .0000    .0000
/
\endpicture}

\def\gluprl{
\beginpicture
\setcoordinatesystem units <\the\scale,\the\scale> point at 0 0
\setplotarea x from 0 to 0, y from 0 to 0
\setplotsymbol ({.})
\setquadratic
\plot
   .0000    .0000    -.3717   1.7852     .5635   3.3646    2.3981   3.4218
  3.8036   2.1987    3.8805   1.2173    3.0580   1.7581    2.6646   3.5793
  3.5997   5.1587    5.4343   5.2160    6.8398   3.9928    6.9168   3.0114
  6.0942   3.5522    5.7008   5.3735    6.6360   6.9529    8.4706   7.0101
  9.8761   5.7870    9.9530   4.8056    9.1305   5.3464    8.7371   7.1676
  9.6722   8.7470   11.5068   8.8043   12.9123   7.5811   12.9893   6.5997
 12.1667   7.1405   11.7733   8.9618   12.7085  10.5412   14.5431  10.5984
 15.9486   9.3753   16.0255   8.3939   15.2030   8.9347   14.8096  10.7559
 15.7447  12.3353   17.5793  12.3926   18.9848  11.1694   19.0618  10.1880
 18.2392  10.7288   17.8458  12.5501   18.7810  14.1295   20.6156  14.1867
 22.0000  13.0000
/
\endpicture}

\def\phohor{
\beginpicture
\setcoordinatesystem units <\the\scale,\the\scale> point at 0 0
\setplotarea x from 0 to 0, y from 0 to 0
\setplotsymbol ({.})
\setquadratic
\plot
   .0000    .0000      .4375    .7071      .8750   1.0000     1.3125    .7071
  1.7500    .0000     2.1875   -.7071     2.6250  -1.0000     3.0625   -.7071
  3.5000    .0000     3.9375    .7071     4.3750   1.0000     4.8125    .7071
  5.2500    .0000     5.6875   -.7071     6.1250  -1.0000     6.5625   -.7071
  7.0000    .0000     7.4375    .7071     7.8750   1.0000     8.3125    .7071
  8.7500    .0000     9.1875   -.7071     9.6250  -1.0000    10.0625   -.7071
 10.5000    .0000    10.9375    .7071    11.3750   1.0000    11.8125    .7071
 12.2500    .0000    12.6875   -.7071    13.1250  -1.0000    13.5625   -.7071
 14.0000    .0000
/
\endpicture}

\def\phover{
\beginpicture
\setcoordinatesystem units <\the\scale,\the\scale> point at 0 0
\setplotarea x from 0 to 0, y from 0 to 0
\setplotsymbol ({.})
\setquadratic
\plot
   .0000    .0000     -.7071    .3750    -1.0000    .7500     -.7071   1.1250
   .0000   1.5000      .7071   1.8750     1.0000   2.2500      .7071   2.6250
   .0000   3.0000     -.7071   3.3750    -1.0000   3.7500     -.7071   4.1250
   .0000   4.5000      .7071   4.8750     1.0000   5.2500      .7071   5.6250
   .0000   6.0000     -.7071   6.3750    -1.0000   6.7500     -.7071   7.1250
   .0000   7.5000      .7071   7.8750     1.0000   8.2500      .7071   8.6250
   .0000   9.0000     -.7071   9.3750    -1.0000   9.7500     -.7071  10.1250
   .0000  10.5000      .7071  10.8750     1.0000  11.2500      .7071  11.6250
   .0000  12.0000
/
\endpicture}

\def\phoupl{
\beginpicture
\setcoordinatesystem units <\the\scale,\the\scale> point at 0 0
\setplotarea x from 0 to 0, y from 0 to 0
\setplotsymbol ({.})
\setquadratic
\plot
-10.0000  10.0000    -9.1875  10.1875    -8.6679  10.0821    -8.5625   9.5625
 -8.7500   8.7500    -8.9375   7.9375    -8.8321   7.4179    -8.3125   7.3125
 -7.5000   7.5000    -6.6875   7.6875    -6.1679   7.5821    -6.0625   7.0625
 -6.2500   6.2500    -6.4375   5.4375    -6.3321   4.9179    -5.8125   4.8125
 -5.0000   5.0000    -4.1875   5.1875    -3.6679   5.0821    -3.5625   4.5625
 -3.7500   3.7500    -3.9375   2.9375    -3.8321   2.4179    -3.3125   2.3125
 -2.5000   2.5000    -1.6875   2.6875    -1.1679   2.5821    -1.0625   2.0625
 -1.2500   1.2500    -1.4375    .4375    -1.3321   -.0821     -.8125   -.1875
   .0000    .0000
/
\endpicture}

\def\phoupr{
\beginpicture
\setcoordinatesystem units <\the\scale,\the\scale> point at 0 0
\setplotarea x from 0 to 0, y from 0 to 0
\setplotsymbol ({.})
\setquadratic
\plot
   .0000    .0000     -.1875    .8125     -.0821   1.3321      .4375   1.4375
  1.2500   1.2500     2.0625   1.0625     2.5821   1.1679     2.6875   1.6875
  2.5000   2.5000     2.3125   3.3125     2.4179   3.8321     2.9375   3.9375
  3.7500   3.7500     4.5625   3.5625     5.0821   3.6679     5.1875   4.1875
  5.0000   5.0000     4.8125   5.8125     4.9179   6.3321     5.4375   6.4375
  6.2500   6.2500     7.0625   6.0625     7.5821   6.1679     7.6875   6.6875
  7.5000   7.5000     7.3125   8.3125     7.4179   8.8321     7.9375   8.9375
  8.7500   8.7500     9.5625   8.5625    10.0821   8.6679    10.1875   9.1875
 10.0000  10.0000
/
\endpicture}

\def\phodol{
\beginpicture
\setcoordinatesystem units <\the\scale,\the\scale> point at 0 0
\setplotarea x from 0 to 0, y from 0 to 0
\setplotsymbol ({.})
\setquadratic
\plot
-10.0000 -10.0000   -10.1875  -9.1875   -10.0821  -8.6679    -9.5625  -8.5625
 -8.7500  -8.7500    -7.9375  -8.9375    -7.4179  -8.8321    -7.3125  -8.3125
 -7.5000  -7.5000    -7.6875  -6.6875    -7.5821  -6.1679    -7.0625  -6.0625
 -6.2500  -6.2500    -5.4375  -6.4375    -4.9179  -6.3321    -4.8125  -5.8125
 -5.0000  -5.0000    -5.1875  -4.1875    -5.0821  -3.6679    -4.5625  -3.5625
 -3.7500  -3.7500    -2.9375  -3.9375    -2.4179  -3.8321    -2.3125  -3.3125
 -2.5000  -2.5000    -2.6875  -1.6875    -2.5821  -1.1679    -2.0625  -1.0625
 -1.2500  -1.2500     -.4375  -1.4375      .0821  -1.3321      .1875   -.8125
   .0000    .0000
/
\endpicture}

\def\phodor{
\beginpicture
\setcoordinatesystem units <\the\scale,\the\scale> point at 0 0
\setplotarea x from 0 to 0, y from 0 to 0
\setplotsymbol ({.})
\setquadratic
\plot
   .0000    .0000      .8125    .1875     1.3321    .0821     1.4375   -.4375
  1.2500  -1.2500     1.0625  -2.0625     1.1679  -2.5821     1.6875  -2.6875
  2.5000  -2.5000     3.3125  -2.3125     3.8321  -2.4179     3.9375  -2.9375
  3.7500  -3.7500     3.5625  -4.5625     3.6679  -5.0821     4.1875  -5.1875
  5.0000  -5.0000     5.8125  -4.8125     6.3321  -4.9179     6.4375  -5.4375
  6.2500  -6.2500     6.0625  -7.0625     6.1679  -7.5821     6.6875  -7.6875
  7.5000  -7.5000     8.3125  -7.3125     8.8321  -7.4179     8.9375  -7.9375
  8.7500  -8.7500     8.5625  -9.5625     8.6679 -10.0821     9.1875 -10.1875
 10.0000 -10.0000
/
\endpicture}

\def\phouprl{
\beginpicture
\setcoordinatesystem units <\the\scale,\the\scale> point at 0 0
\setplotarea x from 0 to 0, y from 0 to 0
\setplotsymbol ({.})
\setquadratic
\plot
   .0000    .0000      .1252    .9121      .4700   1.4730     1.1252   1.5371
  2.0000   1.2500     2.8748    .9629     3.5300   1.0270     3.8748   1.5879
  4.0000   2.5000     4.1252   3.4121     4.4700   3.9730     5.1252   4.0371
  6.0000   3.7500     6.8748   3.4629     7.5300   3.5270     7.8748   4.0879
  8.0000   5.0000     8.1252   5.9121     8.4700   6.4730     9.1252   6.5371
 10.0000   6.2500    10.8748   5.9629    11.5300   6.0270    11.8748   6.5879
 12.0000   7.5000    12.1252   8.4121    12.4700   8.9730    13.1252   9.0371
 14.0000   8.7500    14.8748   8.4629    15.5300   8.5270    15.8748   9.0879
 16.0000  10.0000    16.1252  10.9121    16.4700  11.4730    17.1252  11.5371
 18.0000  11.2500    18.8748  10.9629    19.5300  11.0270    19.8748  11.5879
 20.0000  12.5000    20.1252  13.4121    20.4700  13.9730    21.1252  14.0371
 22.0000  13.7500    22.8748  13.4629    23.5300  13.5270    23.8748  14.0879
 24.0000  15.0000
/
\endpicture}

\def\phoupll{
\beginpicture
\setcoordinatesystem units <\the\scale,\the\scale> point at 0 0
\setplotarea x from 0 to 0, y from 0 to 0
\setplotsymbol ({.})
\setquadratic
\plot
-24.0000  15.0000   -23.1252  15.2871   -22.4700  15.2230   -22.1252  14.6621
-22.0000  13.7500   -21.8748  12.8379   -21.5300  12.2770   -20.8748  12.2129
-20.0000  12.5000   -19.1252  12.7871   -18.4700  12.7230   -18.1252  12.1621
-18.0000  11.2500   -17.8748  10.3379   -17.5300   9.7770   -16.8748   9.7129
-16.0000  10.0000   -15.1252  10.2871   -14.4700  10.2230   -14.1252   9.6621
-14.0000   8.7500   -13.8748   7.8379   -13.5300   7.2770   -12.8748   7.2129
-12.0000   7.5000   -11.1252   7.7871   -10.4700   7.7230   -10.1252   7.1621
-10.0000   6.2500    -9.8748   5.3379    -9.5300   4.7770    -8.8748   4.7129
 -8.0000   5.0000    -7.1252   5.2871    -6.4700   5.2230    -6.1252   4.6621
 -6.0000   3.7500    -5.8748   2.8379    -5.5300   2.2770    -4.8748   2.2129
 -4.0000   2.5000    -3.1252   2.7871    -2.4700   2.7230    -2.1252   2.1621
 -2.0000   1.2500    -1.8748    .3379    -1.5300   -.2230     -.8748   -.2871
   .0000    .0000
/
\endpicture}

\def\arhor{
\beginpicture
\setcoordinatesystem units <\the\scale,\the\scale> point at 0 0
\setplotarea x from 0 to 0, y from 0 to 0
\setplotsymbol ({.})
\arrow <1mm> [1,2] from 0 0  to  4 0
\endpicture}

\def\arver{
\beginpicture
\setcoordinatesystem units <\the\scale,\the\scale> point at 0 0
\setplotarea x from 0 to 0, y from 0 to 0
\setplotsymbol ({.})
\arrow <1mm> [1,2] from 0 4  to  0 0
\endpicture}

\def\arupr{
\beginpicture
\setcoordinatesystem units <\the\scale,\the\scale> point at 0 0
\setplotarea x from 0 to 0, y from 0 to 0
\setplotsymbol ({.})
\arrow <1mm> [1,2] from 0 0  to  3 3
\endpicture}

\def\arupl{
\beginpicture
\setcoordinatesystem units <\the\scale,\the\scale> point at 0 0
\setplotarea x from 0 to 0, y from 0 to 0
\setplotsymbol ({.})
\arrow <1mm> [1,2] from 0 0  to  -3 3
\endpicture}

\def\ardor{
\beginpicture
\setcoordinatesystem units <\the\scale,\the\scale> point at 0 0
\setplotarea x from 0 to 0, y from 0 to 0
\setplotsymbol ({.})
\arrow <1mm> [1,2] from 0 0  to  3 -3
\endpicture}

\def\ardol{
\beginpicture
\setcoordinatesystem units <\the\scale,\the\scale> point at 0 0
\setplotarea x from 0 to 0, y from 0 to 0
\setplotsymbol ({.})
\arrow <1mm> [1,2] from 0 0  to  -3 -3
\endpicture}

\def\schor{
\beginpicture
\setcoordinatesystem units <\the\scale,\the\scale> point at 0 0
\setplotarea x from 0 to 0, y from 0 to 0
\setplotsymbol ({.})
\setdashes
\putrule from 0 0  to 24  0
\setlinear
\plot 0 0   24 0
/
\endpicture}

\def\scver{
\beginpicture
\setcoordinatesystem units <\the\scale,\the\scale> point at 0 0
\setplotarea x from 0 to 0, y from 0 to 0
\setplotsymbol ({.})
\setdashes
\putrule from 0 0  to  0  12
\setlinear
\plot 0 0   0  12
/
\endpicture}

\def\scupr{
\beginpicture
\setcoordinatesystem units <\the\scale,\the\scale> point at 0 0
\setplotarea x from 0 to 0, y from 0 to 0
\setplotsymbol ({.})
\setdashes
\putrule from 0 0  to 10  10
\setlinear
\plot 0 0  10 10
/
\endpicture}

\def\scupl{
\beginpicture
\setcoordinatesystem units <\the\scale,\the\scale> point at 0 0
\setplotarea x from 0 to 0, y from 0 to 0
\setplotsymbol ({.})
\setdashes
\putrule from 0 0  to -10 10
\setlinear
\plot 0 0  -10 10
/
\endpicture}

\def\scdor{
\beginpicture
\setcoordinatesystem units <\the\scale,\the\scale> point at 0 0
\setplotarea x from 0 to 0, y from 0 to 0
\setplotsymbol ({.})
\setdashes
\putrule from 0 0  to 10  -10
\setlinear
\plot 0 0  10 -10
/
\endpicture}

\def\scdol{
\beginpicture
\setcoordinatesystem units <\the\scale,\the\scale> point at 0 0
\setplotarea x from 0 to 0, y from 0 to 0
\setplotsymbol ({.})
\setdashes
\putrule from 0 0  to -10 -10
\setlinear
\plot 0 0  -10 -10
/
\endpicture}

\def\fehor{
\beginpicture
\setcoordinatesystem units <\the\scale,\the\scale> point at 0 0
\setplotarea x from 0 to 0, y from 0 to 0
\setsolid
\putrule from 0  0   to 24  0
\setlinear
\plot 0 0   24 0
/
\endpicture}

\def\fehors{
\beginpicture
\setcoordinatesystem units <\the\scale,\the\scale> point at 0 0
\setplotarea x from 0 to 0, y from 0 to 0
\setsolid
\putrule from 0  0   to 14  0
\setlinear
\plot 0 0   14 0
/
\endpicture}

\def\fever{
\beginpicture
\setcoordinatesystem units <\the\scale,\the\scale> point at 0 0
\setplotarea x from 0 to 0, y from 0 to 0
\setplotsymbol ({.})
\setsolid
\putrule from 0 0  to 0 12
\setlinear
\plot 0 0  0 12
/
\endpicture}

\def\feupr{
\beginpicture
\setcoordinatesystem units <\the\scale,\the\scale> point at 0 0
\setplotarea x from 0 to 0, y from 0 to 0
\setplotsymbol ({.})
\setsolid
\putrule from 0 0  to 10 10
\setlinear
\plot 0 0  10 10
/
\endpicture}

\def\feupl{
\beginpicture
\setcoordinatesystem units <\the\scale,\the\scale> point at 0 0
\setplotarea x from 0 to 0, y from 0 to 0
\setplotsymbol ({.})
\setsolid
\putrule from 0 0  to -10 10
\setlinear
\plot 0 0  -10 10
/
\endpicture}

\def\fedor{
\beginpicture
\setcoordinatesystem units <\the\scale,\the\scale> point at 0 0
\setplotarea x from 0 to 0, y from 0 to 0
\setplotsymbol ({.})
\setsolid
\putrule from 0 0  to 10 -10
\setlinear
\plot 0 0  10 -10
/
\endpicture}

\def\fedol{
\beginpicture
\setcoordinatesystem units <\the\scale,\the\scale> point at 0 0
\setplotarea x from 0 to 0, y from 0 to 0
\setplotsymbol ({.})
\setsolid
\putrule from 0 0  to -10 -10
\setlinear
\plot 0 0  -10 -10
/
\endpicture}


\section*{Appendix}
\label{sec:appendix}

In this appendix, we present all the Feynman rules for propagators
and vertices needed in the present work. The Landau gauge is chosen
for the photon propagator and the de Donder gauge for the graviton
propagator. A dashed line is for a scalar and a directed solid
line for a fermion. A vector boson is denoted by a wiggly line
and a graviton by a curly line.

\beginpicture
\setcoordinatesystem units <\the\scale,\the\scale> point at 0 0
\setplotarea x from -120 to 100, y from -115 to 150
\setplotsymbol ({.})
\put{{\bf\underline{Feynman Rules}}} at 0 95
\put{\stack [l]{$\bullet$ Scalar propagator:}} [l] at -95 60
\put{\schor}[l]  at -10 60
\put{$p$}  at 4 55
\put{$\frac{i}{p^2-m^2}$}  at 80  60
\put{\stack[l]{$\bullet$ Fermion propagator:}} [l]  at -95 40
\put{\fehor}[l]  at -10 40
\put{\arhor}[l]  at 0 40
\put{$p$}  at 4 35
\put{$\frac{i}{\not{p}-m}$}  at 80 40
\put{\stack [l]{$\bullet$ $W$ boson propagator:}} [l] at -95 20
\put{\phohor}[l] at -10 20
\put{\phohor} at 1.5 20
\put{$\mu$} at -13 20
\put{$\nu$} at 23 20
\put{$p$} at 4 15
\put{$\frac{-i(\eta^{\mu\nu}-\frac{p^\mu p^\nu}{m^2})}{p^2-m^2}$}
    at 80 20
\put{\stack [l]{$\bullet$ Photon propagator:}} [l] at -95 0
\put{\phohor}[l] at -10 0
\put{\phohor} at 1.5 0
\put{$\mu$} at -13 0
\put{$\nu$} at 23 0
\put{$p$} at 4 -5
\put{$\frac{-i\eta^{\mu\nu}}{p^2}$} at 80 0
\put{\stack [l]{$\bullet$ Graviton propagator:}} [l] at -95 -20
\put{\glhor}[l] at -10  -20
\put{\glhor} at 1.5 -20
\put{$\mu\nu$} at -13 -20
\put{$\alpha\beta$} at 23 -20
\put{$p$} at 4 -25
\put{$\frac{i}{2}\frac{\eta^{\mu\alpha}\eta^{\nu\beta}
	+\eta^{\mu\beta}\eta^{\nu\alpha}
	-\eta^{\mu\nu}\eta^{\alpha\beta}}{p^2}$} at 80 -20

\put{\stack [l]{$\bullet$ $ss\gamma$ vertex:}} [l] at -95 -45
\put{\phover}[l] at 3  -50
\put{\scdol}[l] at 1 -50
\put{\scdor}[l] at 1 -50
\put{$\mu$} at 9  -37
\put{$\arupr$} at -6 -55
\put{$\ardor$} at  12  -52
\put{$p_1$} at -10 -61
\put{$p_2$} at 21 -61
\put{$-ie(p_1+p_2)_\mu$} at 80 -50

\put{\stack [l]{$\bullet$ $ss\gamma\gamma$ vertex:}} [l] at -95 -80
\put{\phoupl}[l] at 3  -85
\put{\phoupr}[l] at 3  -85
\put{\scdol}[l] at 1 -85
\put{\scdor}[l] at 1 -85
\put{$\mu$} at -10  -75
\put{$\nu$} at 21  -75
\put{$2ie^2\eta_{\mu\nu}$} at 80 -80

\put{\stack [l]{$\bullet$ $ff\gamma$ vertex:}} [l] at -95 -115
\put{\phover}[l] at 3  -120
\put{\fedol}[l] at 1 -120
\put{\fedor}[l] at 1 -120
\put{$\mu$} at 9  -107
\put{$-ie\gamma_\mu$} at 80 -115
\put{$\arupr$} at -3 -128
\put{$\ardor$} at  8  -123
\endpicture

\beginpicture
\setcoordinatesystem units <\the\scale,\the\scale> point at 0 0
\setplotarea x from 0 to 100, y from -100 to 150
\setplotsymbol ({.})
\put{\stack [l]{$\bullet$ $WW\gamma$ vertex:}} [l] at -95 130
\put{\phover}[l] at 3 130
\put{\phodol}[l] at 3 129
\put{\phodor}[l] at 3 129
\put{$\mu$} at 0  143
\put{\arver} at  0 134
\put{$k$} at 11  137
\put{\arupr} at -8 123
\put{\arupl} at  16 124
\put{$p_1$} at -12 121
\put{$p_2$} at 22 121
\put{$W^-_\alpha$} at -11 113
\put{$W^+_\beta$} at 23 113
\put{\stack {$-ie[(p_1-p_2)_\mu \eta_{\alpha\beta}$,
     $\hskip 6mm  +(p_2-k)_\alpha \eta_{\mu\beta}$,
     $\hskip 6mm  +(k-p_1)_\beta \eta_{\alpha\mu}]$}} at 80 130

\put{\stack [l]{$\bullet$ $WW\gamma\gamma$ vertex:}} [l] at -95 90
\put{\phoupl}[l] at 3 90
\put{\phoupr}[l] at 3 90
\put{\phodol}[l] at 3 90
\put{\phodor}[l] at 3 90
\put{$\mu$} at -10 97
\put{$\nu$} at 23  97
\put{$W^-_\alpha$} at -11 74
\put{$W^+_\beta$} at 23 74
\put{$-ie^2(2\eta_{\mu\nu}\eta_{\alpha\beta}
	-\eta_{\mu\alpha}\eta_{\nu\beta}
        -\eta_{\mu\beta}\eta_{\nu\alpha})$} at 80 90

\put{\stack [l]{$\bullet$ $ssg$ vertex:}} [l] at -95 50
\put{\glver}[l] at 5 50
\put{\scdol}[l] at 3 50
\put{\scdor}[l] at 3 50
\put{$\mu\nu$} at 14  59
\put{\arupr} at -5  45
\put{\ardor} at  14 48
\put{$p_1$} at -8 40
\put{$p_2$} at 22  40
\put{\stack {$\frac{i}{2}\kappa[\eta_{\mu\nu}(p_1 \cdot p_2-m^2)$,
           $\hskip 4mm -p_{1\mu} p_{2\nu}
            -p_{1\nu} p_{2\mu}]$}} at 80 50

\put{\stack [l]{$\bullet$ $ssgg$ vertex:}} [l] at -95 20
\put{\glupl}[l] at -60  -5
\put{\glupr}[l] at -60  -5
\put{\scdol}[l] at  -62 -5
\put{\scdor}[l] at  -62 -5
\put{$\mu\nu$} at -77  5
\put{$\lambda\kappa$} at -40  5
\put{\arupr} at -72 -12
\put{\ardor} at -51 -8
\put{$p_1$} at -75 -15
\put{$p_2$} at -41  -15
\put{\stack{$\frac{i}{4} \kappa^2
        [(\eta_{\mu\nu}\eta_{\lambda\kappa}
        -\eta_{\mu\lambda}\eta_{\nu\kappa}
        -\eta_{\mu\kappa}\eta_{\nu\lambda})(p_1 \cdot p_2-m^2)$ ,
 $\hskip 4mm +\eta_{\nu\lambda} (p_{1\mu} p_{2\kappa}
   +p_{1\kappa} p_{2\mu})
   +\eta_{\mu\kappa}(p_{1\lambda}p_{2\nu}+p_{1\nu} p_{2\lambda})$,
 $\hskip 4mm +\eta_{\mu\lambda}(p_{1\nu} p_{2\kappa}
   + p_{1\kappa} p_{2\nu})
   +\eta_{\nu\kappa}(p_{1\lambda} p_{2\mu}
   + p_{1\mu} p_{2\lambda})$ ,
  $\hskip 4mm -\eta_{\mu\nu}(p_{1\lambda}p_{2\kappa}
   +p_{1\kappa}p_{2\lambda})
   -\eta_{\lambda\kappa} (p_{1\mu} p_{2\nu}
   + p_{1\nu} p_{2\mu})]$}}
      at 50 5

\put{\stack [l]{$\bullet$ $ffg$ vertex:}} [l] at -95 -40
\put{\glver}[l] at 3  -40
\put{\fedol}[l] at 1 -40
\put{\fedor}[l] at 1 -40
\put{$\mu\nu$} at 11  -29
\put{\arupr} at -8 -47
\put{\ardor} at 12 -42
\put{\arupr} at -4 -48
\put{\ardor} at 7 -43
\put{$p_1$} at -11 -50
\put{$p_2$} at 21  -50
\put{ \stack{$ \frac{i}{8} \kappa[2\eta_{\mu\nu}
	 (\not{p_1}+\not{p_2}-2m)$,
	 $-(p_1 +p_2)_\mu\gamma_\nu
          -\gamma_\mu(p_1+p_2)_\nu ]$}} at 80 -40

\put{\stack [l]{$\bullet$ $ffgg$ vertex:}} [l] at -95 -80
\put{\glupl}[l] at 3  -85
\put{\glupr}[l] at 3  -85
\put{\fedol}[l] at 1 -85
\put{\fedor}[l] at 1 -85
\put{$\mu\nu$} at -11  -75
\put{$\lambda\kappa$} at 21  -75
\put{\arupr} at -8 -91
\put{\ardor} at 11 -87
\put{\arupr} at -4 -93
\put{\ardor} at 7 -88
\put{$p_1$} at -10 -95
\put{$p_2$} at 21 -95
\put{\stack {$i\kappa^2 [\rm Sym][\rm Per]
     [\frac{3}{16}\eta_{\mu\lambda}\gamma_\kappa (p_1+p_2)_\nu$,
   $+\frac{1}{16}\eta_{\mu\lambda}k_{2\nu}\gamma_\kappa
    +\frac{1}{32}\eta_{\mu\lambda} \not\!{k_2}
    (\gamma_\kappa \gamma_\nu-\gamma_\nu \gamma_\kappa)]$ }}
   at 80 -80
\put{\stack{Here $[{\rm Sym}]$ represents symetrization between
      $\mu$ and $\nu$
       and between $\lambda$ and $\kappa$,,
       while the symbol $[{\rm Per}]$ indicates permutation
       among both $(k_1\mu\nu)$ and
       $(k_2\lambda\kappa)$.}} at 5 -120

\endpicture

\beginpicture
\setcoordinatesystem units <\the\scale,\the\scale> point at 0 0
\setplotarea x from 0 to 100, y from -150 to 150
\setplotsymbol ({.})
\put{\stack [l]{$\bullet$ $WWg$ vertex:}} [l] at -95 142
\put{\glver}[l] at -60 115
\put{\phodol}[l] at -60 115
\put{\phodor}[l] at -60 115
\put{$\lambda\kappa$} at -50  127
\put{\arupr}[l] at -71 111
\put{\ardor}[l] at -53 115
\put{$p_1$} at -73 107
\put{$p_2$} at -43 107
\put{$W^-_\mu$} at -75 99
\put{$W^+_\nu$} at -39 99
\put{\stack [l] {$ -\frac{i}{2} \kappa
         [\eta_{\lambda\kappa}\eta_{\mu\nu}(p_1 \cdot p_2-m^2)
          -\eta_{\lambda\kappa}p_{1\nu}p_{2\mu} $,
        $\hskip8mm  +\eta_{\kappa\mu}p_{1\nu}p_{2\lambda}
         -\eta_{\mu\nu}p_{1\kappa}p_{2\lambda}
         +\eta_{\lambda\nu}p_{1\kappa}p_{2\mu}$,
        $\hskip8mm  -\eta_{\kappa\mu}\eta_{\lambda\nu}
         (p_1 \cdot p_2-m^2)
         +\eta_{\kappa\nu}p_{1\lambda}p_{2\mu}$,
        $\hskip8mm  -\eta_{\mu\nu}p_{1\lambda}p_{2\kappa}
         +\eta_{\lambda\mu}p_{1\nu}p_{2\kappa}$,
	 $\hskip8mm -\eta_{\kappa\nu}\eta_{\lambda\mu}
	 (p_1 \cdot p_2-m^2)]$}} [lt] at 10 140

\put{\stack [l]{$\bullet$ $WWgg$ vertex:}} [l] at -95 75
\put{\glupl}[l] at -60 40
\put{\glupr}[l] at -60 40
\put{\phodol}[l] at -60 40
\put{\phodor}[l] at -60 40
\put{$\lambda\kappa$} at -76 50
\put{$\rho\sigma$} at -40 50
\put{\arupr}[l] at -71 35
\put{\ardor}[l] at -52 39
\put{$p_1$} at -73 32
\put{$p_2$} at -43 32
\put{$W^-_\mu$} at -75 24
\put{$W^+_\nu$} at -39 24
\put{\stack [l] {$-\frac{i}{4} \kappa^2
      [(\eta_{\lambda\kappa}\eta_{\rho\sigma}
       -2\eta_{\lambda\rho}\eta_{\kappa\sigma})
      (\eta_{\mu\nu}(p_1 \cdot p_2-m^2)-p_{1\nu}p_{2\mu})$,
    $\hskip8mm -\eta_{\lambda\kappa}
      (T_{\mu\nu\rho\sigma}+T_{\mu\nu\sigma\rho})
     -\eta_{\rho\sigma}(T_{\mu\nu\lambda\kappa}
     +T_{\mu\nu\kappa\lambda})$,
    $\hskip8mm +2\eta_{\kappa\rho}
     (T_{\mu\nu\lambda\sigma}+T_{\mu\nu\sigma\lambda})
     +2\eta_{\lambda\sigma}(T_{\mu\nu\rho\kappa}
     +T_{\mu\nu\kappa\rho})
     $,$\hskip10mm +2(\eta_{\rho\mu}\eta_{\nu\sigma}
      p_{1\lambda}p_{2\kappa}
     -\eta_{\mu\lambda}\eta_{\nu\sigma}p_{1\rho}p_{2\kappa}
     -\eta_{\mu\rho}\eta_{\nu\kappa}p_{1\lambda}p_{2\sigma}
     $,$\hskip15mm+ \eta_{\mu\sigma}\eta_{\nu\rho}p_{1\kappa}
       p_{2\lambda}
      -\eta_{\mu\kappa}\eta_{\nu\rho}p_{1\sigma}p_{2\lambda}
      -\eta_{\mu\sigma}\eta_{\nu\lambda}p_{1\kappa}p_{2\rho}
      $,$\hskip15mm +\eta_{\mu\lambda}\eta_{\nu\kappa}
       p_{1\rho}p_{2\sigma}
      +\eta_{\mu\kappa}\eta_{\nu\lambda}p_{1\sigma}p_{2\rho})]$,
   where,$ \ \ $,
   $\hskip10mm T_{\mu\nu\rho\sigma}
       =\eta_{\mu\nu}p_{1\rho}p_{2\sigma}
       -\eta_{\mu\rho}p_{1\nu}p_{2\sigma}$,$\hskip15mm
       -\eta_{\nu\sigma}p_{1\rho}p_{2\mu}
       +\eta_{\mu\rho}\eta_{\nu\sigma}(p_1\cdot p_2-m^2)$}}
     [lt] at -4 85

\put{\stack [l]{$\bullet$ $ss\gamma g$ vertex:}} [l] at -95 -10
\put{\glupl}[l] at 5  -10
\put{\phoupr}[l] at 5 -10
\put{\scdol}[l] at 3 -10
\put{\scdor}[l] at 3 -10
\put{$\mu\nu$} at -12  0
\put{$\alpha$} at 23  0
\put{\arupr}[l] at -8 -16
\put{\ardor}[l] at 14 -13
\put{$p_1$} at -9 -20
\put{$p_2$} at 22  -21
\put{\stack [l]{$\frac{i}{2}
            e\kappa[\eta_{\mu\nu}{(p_1+p_2)}_\alpha$,
             $\hskip6mm -\eta_{\alpha\mu}{(p_1+p_2)}_\nu$,
             $\hskip6mm -\eta_{\alpha\nu}{(p_1+p_2)}_\mu] $}}
           [lt] at 50 0

\put{\stack [l]{$\bullet$ $ff\gamma g$ vertex:}} [l] at -95 -45
\put{\glupl}[l] at 5 -45
\put{\phoupr}[l] at 5 -45
\put{\fedol}[l] at 3 -45
\put{\fedor}[l] at 3 -45
\put{$\mu\nu$} at -12  -33
\put{$\alpha$} at 23  -33
\put{\arupr}[l] at -3 -53
\put{\ardor}[l] at 8 -48
\put{\stack [l] {$-\frac{i}{4}e
                  \kappa[2\eta_{\mu\nu}\gamma_\alpha
                 -\eta_{\alpha\mu}\gamma_\nu
                 -\eta_{\alpha\nu}\gamma_\mu]$}} [lt] at 45 -40

\put{\stack [l]{$\bullet$ $WW\gamma g$ vertex:}} [l] at -95 -75
\put{\glupl}[l] at -60  -105
\put{\phoupr}[l] at -60  -105
\put{\phodol}[l] at  -62 -105
\put{\phodor}[l] at  -62 -105
\put{$\mu\nu$} at -77  -95
\put{$\alpha$} at -42  -94
\put{$W^-_\rho$} at -75 -123
\put{$W^+_\sigma$} at -39 -123
\put{\arupr}[l] at -74 -110
\put{\arupr}[l] at -53 -104
\put{\ardor}[l] at -53 -107
\put{$p_1$} at -75 -115
\put{$p_2$} at -43  -115
\put{$k_2$} at -42  -101
\put{\stack [l] {$\frac{i}{2}e\kappa
   [\eta_{\mu\nu}(\eta_{\rho\sigma} {(p_1+p_2)}_\alpha
                       -\eta_{\alpha\sigma} p_{2\rho}
                       -\eta_{\alpha\rho} p_{1\sigma}) $,
       $\hskip10mm -\eta_{\rho\sigma}({(p_1+p_2)}_\mu
                   \eta_{\nu\alpha}
                   +{(p_1+p_2)}_\nu \eta_{\mu\alpha})$,
        $\hskip10mm -{(p_1+p_2)}_\alpha
                (\eta_{\mu\sigma}\eta_{\nu\rho}
                +\eta_{\nu\sigma}\eta_{\mu\rho})$,
       $\hskip10mm  +\eta_{\alpha\rho}[{(p_1+k_2)}_\mu
                  \eta_{\nu\sigma}
                  +{(p_1+k_2)}_\nu\eta_{\mu\sigma}]$,
       $\hskip10mm +\eta_{\alpha\sigma}[{(p_2-k_2)}_\mu
               \eta_{\nu\rho}
             +{(p_2-k_2)}_\nu\eta_{\mu\rho}]$,
        $\hskip10mm  +{(p_1+k_2)}_\sigma
             (\eta_{\mu\alpha}\eta_{\nu\rho}
            +\eta_{\nu\alpha}\eta_{\mu\rho})$,
       $\hskip10mm +{(p_2-k_2)}_\rho(\eta_{\mu\sigma}
              \eta_{\nu\alpha}
            +\eta_{\nu\sigma}\eta_{\mu\alpha}) ]$}} [lt]
           at 0 -75
\endpicture

\beginpicture
\setcoordinatesystem units <\the\scale,\the\scale> point at 0 0
\setplotarea x from -100 to 100, y from -100 to 150
\setplotsymbol ({.})
\put{\stack [l]{$\bullet$ $ggg$ vertex:}} [l] at -95 142
\put{\glver}[l] at -80 115
\put{\gldol}[l] at -80 115
\put{\gldor}[l] at -80 115
\put{$\mu\alpha$} at -70  128
\put{\arver}[l] at -75 120
\put{\arupr}[l] at -90 111
\put{\arupl}[l] at -67 111
\put{$k_1$} at -86 124
\put{$k_2$} at -94 110
\put{$k_3$} at -60 110
\put{$\nu\beta$} at -95 100
\put{$\sigma\gamma$} at -56 100
\put{\stack [l] {$ i\kappa \ [\rm Sym] [\frac{1}{2}P_6
(k_1 \cdot k_2 \eta_{\mu\alpha}\eta_{\nu\sigma}\eta_{\beta\gamma})
$,$-P_3(k_{1\sigma}k_{2\gamma}\eta_{\mu\nu}\eta_{\alpha\beta})
 -2P_3(k_1 \cdot k_2\eta_{\alpha\nu}
  \eta_{\beta\sigma}\eta_{\gamma\mu})$,$
  +\frac{1}{2}P_3(k_1 \cdot k_2 \eta_{\mu\nu}\eta_{\alpha\beta}
       \eta_{\sigma\gamma})
 +P_6(k_{1\sigma}k_{2\mu} \eta_{\alpha\nu}\eta_{\beta\gamma})$,$
 -\frac{1}{4}P_3(k_1 \cdot k_2 \eta_{\mu\alpha}\eta_{\nu\beta}
         \eta_{\sigma\gamma})]$}} [lt] at -25 140

\put{\stack [l]{$\bullet$ $gggg$ vertex:}} [l] at -95 75
\put{\glupl}[l] at -80 30
\put{\glupr}[l] at -80 30
\put{\gldol}[l] at -80 30
\put{\gldor}[l] at -80 30
\put{$\mu\alpha$} at -96 45
\put{$\nu\beta$} at -60 45
\put{\ardor}[l] at -92 34
\put{\ardol}[l] at -69 34
\put{\arupr}[l] at -92 24
\put{\arupl}[l] at -67 24
\put{$k_1$} at -96 39
\put{$k_2$} at -59 39
\put{$k_4$} at -96  22
\put{$k_3$} at -59  22
\put{$\rho\lambda$} at -96 13
\put{$\sigma\gamma$} at -60 13
\put{\stack [l] {$i \kappa^2 [\rm Sym][
  \frac{1}{4}P_6(k_1 \cdot k_2 \eta_{\mu\nu}\eta_{\alpha\beta}
  \eta_{\sigma\gamma} \eta_{\rho\lambda})$,
  $+2P_{12}(k_{1\sigma}k_{2\gamma}\eta_{\mu\nu}\eta_{\alpha\rho}
   \eta_{\beta\lambda})
   -\frac{1}{2}P_6(k_1 \cdot k_2 \eta_{\mu\nu}\eta_{\alpha\beta}
  \eta_{\sigma\rho} \eta_{\gamma\lambda})$,
 $-4P_{12}(k_{1\sigma}k_{2\mu}\eta_{\alpha\nu}\eta_{\beta\rho}
  \eta_{\gamma\lambda})
+\frac{1}{2}P_{24}(k_{1\sigma}k_{2\mu}\eta_{\alpha\nu}
 \eta_{\beta\gamma}
  \eta_{\rho\lambda})$,
  $-P_{12}(k_1 \cdot k_2 \eta_{\mu\nu}\eta_{\alpha\sigma}
  \eta_{\beta\gamma} \eta_{\rho\lambda})
  +4P_{6}(k_1 \cdot k_2 \eta_{\mu\nu}\eta_{\alpha\sigma}
  \eta_{\beta\rho} \eta_{\gamma\lambda})$,
 $ -P_{12}(k_1 \cdot k_2 \eta_{\mu\sigma}\eta_{\alpha\gamma}
  \eta_{\nu\rho} \eta_{\beta\lambda})
  +2P_{6}(k_1 \cdot k_2 \eta_{\mu\sigma}\eta_{\alpha\rho}
  \eta_{\nu\gamma} \eta_{\beta\lambda})$,
 $-\frac{1}{2}P_{12}(k_{1\sigma}k_{2\gamma}\eta_{\mu\nu}
   \eta_{\alpha\beta}\eta_{\rho\lambda})
  +P_{12}(k_{1\sigma}k_{2\rho}\eta_{\mu\nu}\eta_{\alpha\beta}
   \eta_{\gamma\lambda})$,
  $-2P_{12}(k_{1\sigma}k_{2\rho}\eta_{\mu\nu}\eta_{\alpha\gamma}
   \eta_{\beta\lambda})
  +\frac{1}{4}P_{24}(k_1 \cdot k_2 \eta_{\mu\alpha}
   \eta_{\nu\sigma}
   \eta_{\beta\gamma} \eta_{\rho\lambda})$,
 $ -\frac{1}{2}P_{24}(k_1 \cdot k_2 \eta_{\mu\alpha}
  \eta_{\nu\sigma}
  \eta_{\beta\rho} \eta_{\gamma\lambda})
 +\frac{1}{2}P_{24}(k_{1\beta}k_{2\sigma}\eta_{\mu\alpha}
   \eta_{\nu\rho}\eta_{\gamma\lambda})$,
 $ -P_{6}(k_{1\nu}k_{2\mu}\eta_{\alpha\beta}\eta_{\sigma\rho}
   \eta_{\gamma\lambda})
  -\frac{1}{8}P_{6}(k_1 \cdot k_2 \eta_{\mu\alpha}
  \eta_{\nu\beta}\eta_{\sigma\gamma} \eta_{\rho\lambda})$,
 $ +\frac{1}{2}P_{6}(k_1 \cdot k_2 \eta_{\mu\alpha}
  \eta_{\nu\beta}\eta_{\sigma\rho} \eta_{\gamma\lambda})]$}} [lt]
 at -25 75
\endpicture
\vskip -4cm
\noindent
Here the symbol $[{\rm Sym}]$ means symmetrization between $\mu$
and $\alpha$, between $\nu$ and $\beta$, and between $\sigma$ and
$\gamma$, respectively, for the 3-graviton vertex, or between
$\mu$ and $\alpha$, between $\nu$ and $\beta$, between
$\sigma$ and $\gamma$, and between $\rho$ and $\lambda$,
respectively, for the 4-graviton vertex.
The ${\rm P}$ indicates permutation among $(k_1 \mu \alpha),
(k_2 \nu\beta)$, $(k_3\sigma\gamma)$ for the 3-graviton vertex,
or among $(k_1 \mu\alpha), (k_2 \nu\beta), (k_3 \sigma\gamma)$,
and $(k_4 \rho\lambda)$ for the 4-graviton vertex, and each
subscript in ${\rm P}$ is for the number of independent
permutations. As an example, $P_3(k_1\cdot k_2 \eta_{\mu\nu}
\eta_{\alpha\beta}\eta_{\sigma\gamma})
=(k_1\cdot k_2)\eta_{\mu\nu}\eta_{\alpha\beta}\eta_{\sigma\gamma}
+(k_2\cdot k_3)\eta_{\nu\sigma}\eta_{\beta\gamma}\eta_{\mu\alpha}
+(k_3\cdot k_1)\eta_{\sigma\mu}\eta_{\gamma\alpha}\eta_{\nu\beta}$.

\newpage

\section*{Figure Captions}

\begin{enumerate}
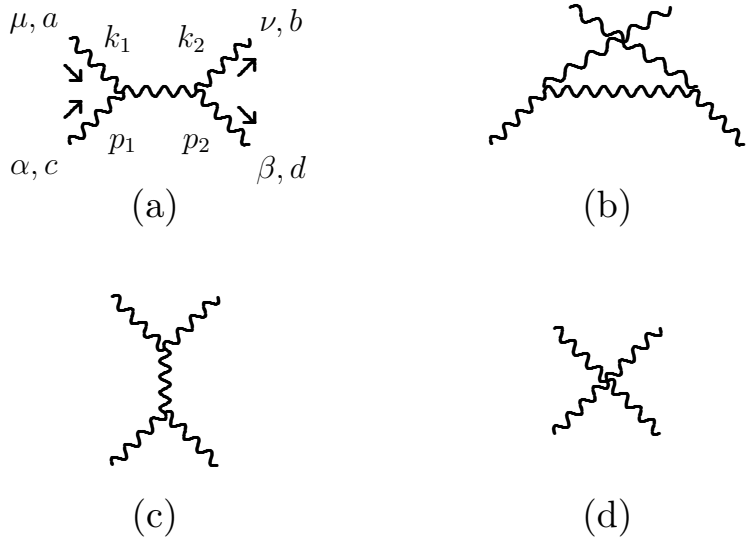
\item[{\bf Fig.~1}]
     Feynman diagrams for the gluon-gluon scattering process
     $G^aG^c\rightarrow  G^bG^d$. The wavy line represents
     a gluon but not a photon.
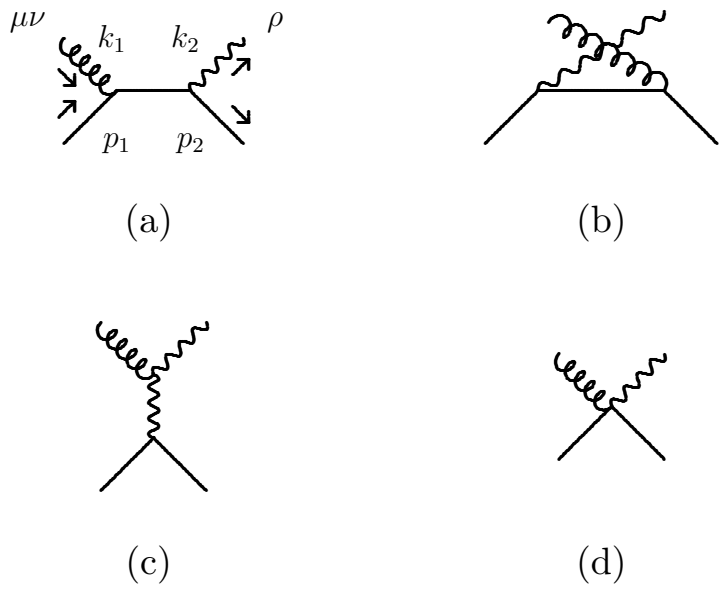
\item[{\bf Fig.~2}]
     Feynman diagrams for the process $gX\rightarrow \gamma X$.
     The curly line is for a graviton and the wavy line for a
     photon.  Here, $X$, represented by a solid line, can be
     a scalar $s$, a fermion $f$, or a vector boson $W$.
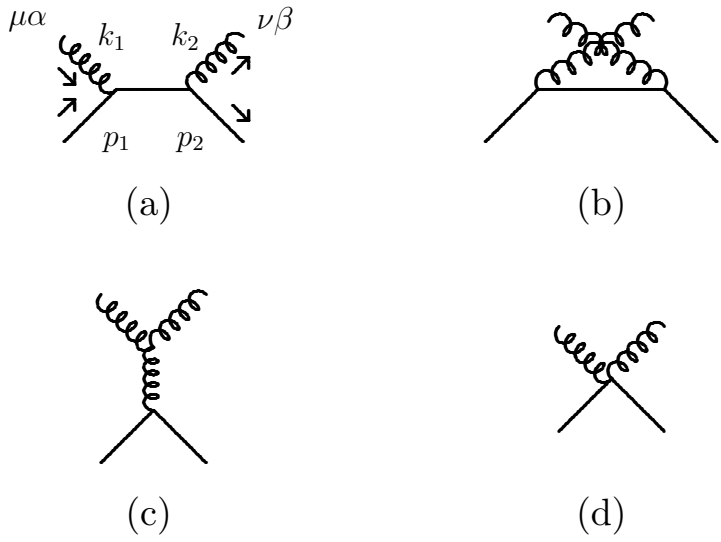
\item[{\bf Fig.~3}]
     Feynman diagrams for the process $gX\rightarrow gX$.
     The curly line is for a graviton.
     $X$,  denoted  by a solid line, can be
     a scalar $s$, a fermion $f$, or a vector boson $W$.
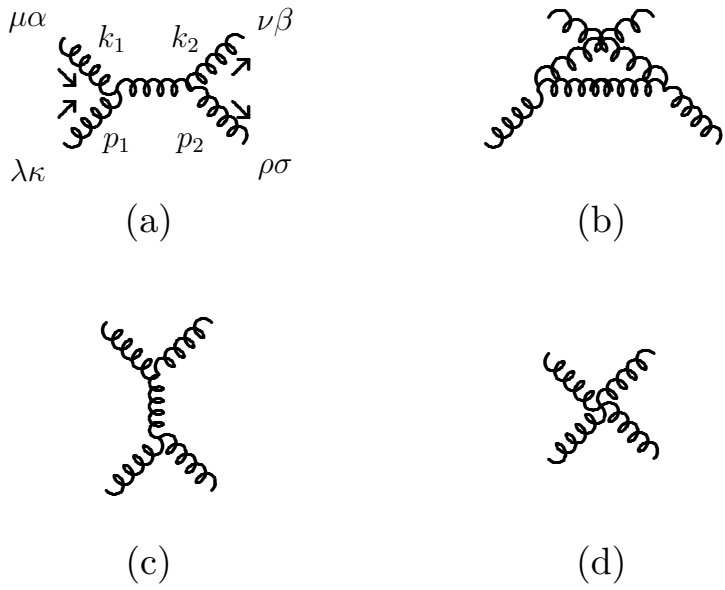
\item[{\bf Fig.~4}]
     Feynman diagrams for the process $gg\rightarrow gg$.
     The curly line is for a graviton.
\end{enumerate}

\newpage
\pagestyle{empty}
\beginpicture
\setcoordinatesystem units <\the\scale,\the\scale> point at 0 0
\setplotarea x from -120 to 0, y from -150 to 200
\setplotsymbol ({.})
\put{\phohor}[l] at -80 185
\put{\phodol}[l] at -80 185
\put{\phoupl}[l] at -80 185
\put{\phodor}[l] at -66 185
\put{\phoupr}[l] at -66 185
\put{$\mu,a$} at -95  198
\put{$\nu,b$} at -48  198
\put{$k_1$} at -79 195
\put{\ardor} at -90 190
\put{$p_1$} at -78 175
\put{\arupr} at -90 180
\put{$k_2$} at -65 195
\put{\arupr} at -57 188
\put{$p_2$} at -64 175
\put{\ardor} at -57 182
\put{$\alpha,c$} at -95 170
\put{$\beta,d$} at -48 170
\put{{\bf\large (a)}} at -72 163

\put{\phohor}[l] at 0 185
\put{\phohor}[l] at 14 185
\put{\phodol}[l] at 0 185
\put{\phoupll}[l] at 29 186
\put{\phodor}[l] at 28 185
\put{\phouprl}[l] at 0 186
\put{{\bf\large (b)}} at 14 163

\put{\phover}[l] at -72 124
\put{\phodol}[l] at -72 124
\put{\phoupl}[l] at -72 136
\put{\phodor}[l] at -72 124
\put{\phoupr}[l] at -72 136
\put{{\bf\large (c)}} at -72 104

\put{\phodol}[l] at 12 130
\put{\phoupl}[l] at 12 130
\put{\phodor}[l] at 12 130
\put{\phoupr}[l] at 12 130
\put{{\bf\large (d)}} at 14 104

\put{{\bf\Large Fig.~1}}
     at -22 75

\put{\fehors}[l] at -80 15
\put{\fedol}[l] at -82 15
\put{\glupl}[l] at -80 15
\put{\fedor}[l] at -68 15
\put{\phoupr}[l] at -66 15
\put{$\mu\nu$} at -95  28
\put{$\rho$} at -48  28
\put{$k_1$} at -79 25
\put{\ardor} at -90 19
\put{$p_1$} at -78 5
\put{\arupr} at -90 10
\put{$k_2$} at -65 25
\put{\arupr} at -57 18
\put{$p_2$} at -64 5
\put{\ardor} at -57 12
\put{{\bf\large (a)}} at -72 -10

\put{\fehor}[l] at 0 15
\put{\fedol}[l] at -2 15
\put{\glupll}[l] at 24 15
\put{\fedor}[l] at 22 15
\put{\phouprl}[l] at 0 15
\put{{\bf\large (b)}} at 14 -10

\put{\phover}[l] at -73 -51
\put{\fedol}[l] at -75 -51
\put{\glupl}[l] at -73 -39
\put{\fedor}[l] at -75 -51
\put{\phoupr}[l] at -73 -39
\put{{\bf\large (c)}} at -72 -75

\put{\fedol}[l] at 12 -45
\put{\glupl}[l] at 14 -45
\put{\fedor}[l] at 12 -45
\put{\phoupr}[l] at 14 -45
\put{{\bf\large (d)}} at 14  -75
\put{{\bf\Large Fig.~2}}  at -22 -105

\endpicture

\beginpicture
\setcoordinatesystem units <\the\scale,\the\scale> point at 0 0
\setplotarea x from -120 to 0, y from -150 to 200
\setplotsymbol ({.})

\put{\fehors}[l] at -80 185
\put{\fedol}[l] at -82 185
\put{\glupl}[l] at -80 185
\put{\fedor}[l] at -68 185
\put{\glupr}[l] at -66 185
\put{$\mu\alpha$} at -95  198
\put{$\nu\beta$} at -48  198
\put{$k_1$} at -79 195
\put{\ardor} at -90 189
\put{$p_1$} at -78 175
\put{\arupr} at -90 180
\put{$k_2$} at -65 195
\put{\arupr} at -57 188
\put{$p_2$} at -64 175
\put{\ardor} at -57 182
\put{{\bf\large (a)}} at -72 163

\put{\fehor}[l] at 0 185
\put{\fedol}[l] at -2 185
\put{\glupll}[l] at 24 185
\put{\fedor}[l] at 22 185
\put{\gluprl}[l] at 0 185
\put{{\bf\large (b)}} at 14 163

\put{\glver}[l] at -73 124
\put{\fedol}[l] at -75 124
\put{\glupl}[l] at -73 136
\put{\fedor}[l] at -75 124
\put{\glupr}[l] at -73 136
\put{{\bf\large (c)}} at -72 104

\put{\fedol}[l] at 12 130
\put{\glupl}[l] at 14 130
\put{\fedor}[l] at 12 130
\put{\glupr}[l] at 14 130
\put{{\bf\large (d)}} at 14 104

\put{{\bf\Large Fig.~3}} at -22 75

\put{\glhor}[l] at -80 15
\put{\gldol}[l] at -80 15
\put{\glupl}[l] at -80 15
\put{\gldor}[l] at -66 15
\put{\glupr}[l] at -66 15
\put{$\mu\alpha$} at -95 28
\put{$\nu\beta$} at -48 28
\put{$\lambda\kappa$} at -95 0
\put{$\rho\sigma$} at -48 0
\put{$k_1$} at -79 25
\put{\ardor} at -90 19
\put{$p_1$} at -78 5
\put{\arupr} at -90 10
\put{$k_2$} at -65 25
\put{\arupr} at -57 18
\put{$p_2$} at -64 5
\put{\ardor} at -57 13
\put{{\bf\large (a)}} at -72 -10

\put{\glhor}[l] at 0 15
\put{\glhor}[l] at 10 15
\put{\gldol}[l] at 0 15
\put{\glupll}[l] at 24 16
\put{\gldor}[l] at 24 15
\put{\gluprl}[l] at 0 16
\put{{\bf\large (b)}} at 14 -10

\put{\glver}[l] at -72 -51
\put{\gldol}[l] at -72 -51
\put{\glupl}[l] at -72 -39
\put{\gldor}[l] at -72 -51
\put{\glupr}[l] at -72 -39
\put{{\bf\large (c)}} at -72 -75

\put{\gldol}[l] at 12 -45
\put{\glupl}[l] at 12 -45
\put{\gldor}[l] at 12 -45
\put{\glupr}[l] at 12 -45
\put{{\bf\large (d)}} at 14 -75

\put{{\bf\Large Fig.~4}}
     at -22 -105

\endpicture
\end{document}